\documentclass[superscriptaddress,aps,prl,reprint,twocolumn,amsmath,amssymb]{revtex4-2}
\usepackage{graphicx}
\usepackage{hyperref}
\usepackage{amssymb}
\usepackage{slashed}
\usepackage{dcolumn}
\usepackage{amsmath}
\usepackage{bm}
\usepackage{colordvi}
\usepackage{algorithm}
\usepackage{algpseudocode}
\usepackage{multirow}
\usepackage{titlesec}
\usepackage{newtxtext,newtxmath}
\usepackage{dsfont}
\usepackage{xcolor}
\usepackage{mathbbol}

\usepackage{hyperref}
\hypersetup{
    colorlinks=true,
    linkcolor=blue,
    citecolor=blue,     
    urlcolor=blue,
}

\allowdisplaybreaks

\usepackage{mathrsfs}
\makeatletter

\newcommand{\Rmnum}[1]{\expandafter\@slowromancap\romannumeral #1@}
\makeatother

\begin{document}

\title{Emergent spin-resolved electronic density waves from strong $d$-wave altermagnetism and pseudogap phenomena from phason fluctuations}

\author{Fei Yang}
\email{fzy5099@psu.edu}

\affiliation{Department of Materials Science and Engineering and Materials Research Institute, The Pennsylvania State University, University Park, PA 16802, USA}

\author{Guo-Dong Zhao}
\affiliation{Department of Materials Science and Engineering and Materials Research Institute, The Pennsylvania State University, University Park, PA 16802, USA}

\author{Binghai Yan}
\affiliation{Department of Physics, The Pennsylvania State University, University Park, PA 16802, USA}
\affiliation{The Center for Theory of Emergent Quantum Matter, The Pennsylvania State University, University Park, PA 16802, USA}

\author{Long-Qing Chen}
\email{lqc3@psu.edu}

\affiliation{Department of Materials Science and Engineering and Materials Research Institute, The Pennsylvania State University, University Park, PA 16802, USA}

\date{\today}

\begin{abstract} 
Metallic $d$-wave altermagnets provide a unique setting in which strong spin-momentum locking can qualitatively reshape electronic instabilities. Here we develop a self-consistent microscopic theory beyond the mean-field approximation for density-wave order in $d$-wave altermagnetic metals. The altermagnetic band structure with strong spin-momentum locking reconstructs the Fermi surface into mutually orthogonal spin-selective quasi-1D sectors, whose strong nesting drives density-wave instabilities in the respective spin sectors. The resulting spin-resolved density-wave orders nevertheless share a common ordering wave vector and give rise to a distinct class of stripe states, accompanied by pronounced gap openings on the nested Fermi sheets. The relative phase between the two spin sectors controls the character of the collective order, allowing charge density wave, spin density wave, and mixed density-wave orders to emerge within the same ordered manifold. As temperature increases from zero, thermal excitation of the emergent phason mode induces strong phase fluctuations that destroy long-range density-wave order at $T_c$, while the single-particle gap remains finite, giving rise to a robust pseudogap regime that persists up to a higher temperature $T_g$.  The theory reveals a density-wave mechanism unique to metallic $d$-wave altermagnets, rooted in strong spin-selective Fermi-surface reconstruction, and establishes a fluctuation-driven route to pseudogap phenomena governed by phason dynamics. Applied to the metallic $d$-wave altermagnet KV$_2$Se$_2$O, our simulations quantitatively reproduce the key spectroscopic features reported in recent experiments, providing a unified microscopic understanding of the underlying phenomena.
\end{abstract}

\maketitle  

{\sl Introduction.---}Over the past few years, altermagnetism has been identified as a distinct form of magnetic order that lies outside the conventional ferromagnetic--antiferromagnetic dichotomy~\cite{vsmejkal2022beyond,vsmejkal2022emerging,bhowal2024ferroically,reichlova2024observation,ding2024large,PhysRevLett.133.156702,zhang2025crystal,PhysRevLett.132.166702,Kessler2024}. In altermagnets, oppositely spin-polarized electronic states reside on symmetry-related sublattices connected by crystalline rotations. As a result, the time-reversal symmetry is broken without generating a macroscopic magnetization~\cite{vsmejkal2022beyond,vsmejkal2022emerging,bhowal2024ferroically}.  A defining hallmark of altermagnetism is the emergence of an even-parity, momentum-dependent spin splitting that remains compatible with inversion symmetry. This splitting originates from symmetry-enforced band degeneracies and typically exhibits higher-angular-momentum form factors, such as $d$-, $g$-, or even $i$-wave structures~\cite{vsmejkal2022beyond,vsmejkal2022emerging}.  This unique symmetry-protected spin splitting in altermagnets gives rise to highly anisotropic, spin-resolved electronic dispersions, making these systems particularly attractive for spintronic applications~\cite{bai2024altermagnetism,weissenhofer2024atomistic,gonzalez2021efficient,PhysRevLett.130.216701,sun2025spin,jungwirth2025altermagnetic} as well as emerging quantum-transport~\cite{feng2022anomalous,liao2024separation,sato2024altermagnetic,ma2021multifunctional,fu2025all} or superconductivity-based~\cite{fukaya2025superconducting,fukaya2025josephson,PhysRevB.111.184515,PhysRevLett.131.076003,PhysRevB.108.054511,dlpb-gfct,rqp1-jtcb} technologies.

\begin{figure}[htb]
  {\includegraphics[width=8.5cm]{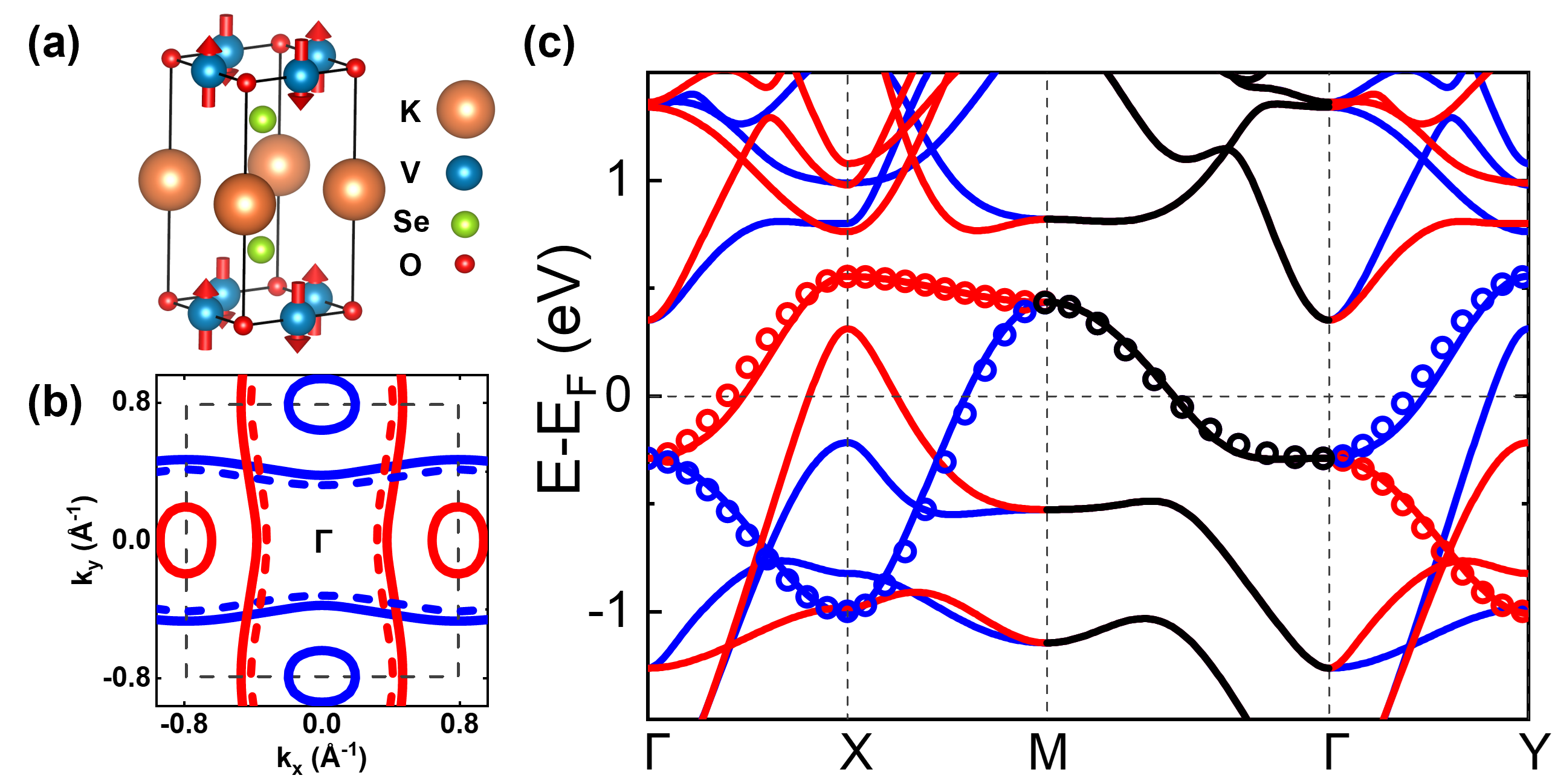}}
\caption{Atomic and electronic structure of KV$_2$Se$_2$O obtained from DFT calculations.
({\bf a}) Crystal unit cell, where the arrows indicate the spin orientations.
({\bf b}) Fermi surface and ({\bf c}) band structure in the $k_z=0$ plane.
Red, blue, and black curves denote spin-up, spin-down, and spin-degenerate bands, respectively.
The dashed curves in ({\bf b}) represent the Fermi surfaces translated by ${\bf Q}=(\pi,\pi)$, and their parallel overlap with the original Fermi surfaces demonstrates a nearly perfect nesting in this system. The metallic electron bands exhibit negligible $k_z$ dispersion (see Sec.~SVIII), allowing the relevant low-energy structure to be effectively described by a 2D altermagnetic tight-binding model.
The circles in ({\bf c}) show the tight-binding band structure, using the same color scheme as the DFT bands.}
\label{figyc1}
\end{figure}

Early realizations of altermagnetism, such as MnTe~\cite{Krempasky2024,PhysRevB.109.115102,PhysRevLett.132.036702} and MnTe$_2$~\cite{Zhu2024}, were predominantly semiconducting. Very recently, KV$_2$Se$_2$O and several other members of the V$_2$Se$_2$O family have been identified as van der Waals layered metallic $d$-wave altermagnets with altermagnetic order persisting up to room temperature~\cite{Jiang2025,bf1n-sxdl,r8nc-dpt8,yang2025altermagnetic,yan2025sdw,Cheng2026}, supported by a combination of spin- and angle-resolved photoemission spectroscopy (ARPES)~\cite{Jiang2025}, nuclear magnetic resonance (NMR)~\cite{Jiang2025}, and density-functional-theory (DFT) calculations~\cite{bf1n-sxdl,r8nc-dpt8,yang2025altermagnetic,yan2025sdw,Cheng2026}. A striking feature revealed by these studies is a strong momentum-dependent spin splitting of the metallic electron bands [Fig.~\ref{figyc1}(c)]: for one spin sector, the dispersion exhibits opposite curvatures along two orthogonal crystallographic directions, while the curvature pattern is reversed for the opposite spin sector. This unusual dispersion reconstructs the Fermi surface into open,  spin-selective quasi-one-dimensional (1D) sheets~\cite{EIdiscussion}, as illustrated in Fig.~\ref{figyc1}(b). 

Such quasi-1D Fermi-surface sheets in an itinerant electronic system naturally suggests a  generically enhanced susceptibility to collective ordering. Experimentally, NMR spectra of KV$_2$Se$_2$O~\cite{Jiang2025} reveal a transition from a uniform magnetic response at high temperatures to a distinct spectral splitting upon cooling below $100~$K,  indicating the emergence of an electronically ordered phase. 
Meanwhile, ARPES measurements~\cite{Jiang2025} reveal pronounced gap openings along the $\Gamma$-X and X/Y-M directions of the Brillouin zone. The low-temperature phase is then attributed, {both experimentally and} in several DFT-based studies~\cite{yan2025sdw,Jiang2025,r8nc-dpt8}, to a spin-density-wave (SDW) instability with an ordering vector ${\bf Q}=(\pi,\pi)$. Interestingly, the gap opening was observed to persist up to 230~K~\cite{Jiang2025}, far above the ordering temperature 100~K, and is interpreted as a pseudogap regime whose microscopic origin remains unclear.
 
Theoretically, quasi-1D electronic structures are
generically susceptible to charge-density-wave (CDW)
instabilities~\cite{gruner1985charge,gruner1988dynamics,lee1974conductivity,
peierls1996quantum,PhysRevB.97.224423,PhysRevLett.35.120,PhysRevB.42.6498}.
In particular, as shown in Fig.~\ref{figyc1}(b), the spin-up and spin-down metallic
bands exhibit nearly perfect nesting at ${\bf Q}=(\pi,\pi)$, suggesting
strong density-wave instabilities. The experimentally
observed pseudogap further points to an electronic instability beyond the
single-particle description. In this work, we develop a self-consistent microscopic theory beyond the mean-field approximation and report a distinct density-wave mechanism rooted in the spin-momentum locking of $d$-wave altermagnetism. The spin-selective Fermi-surface reconstruction drives density-wave instabilities in the two spin sectors at a common ordering wave vector, producing pronounced gap openings on the nested Fermi sheets. The relative phase between the two spin sectors provides an internal degree of freedom that controls the character of the collective order, encompassing CDW, SDW, and mixed density-wave states within a unified ordered manifold. Beyond mean field, thermally excited phasons, emerging as collective modes of the electronic density-wave condensate, generate strong phase fluctuations that can melt the density-wave order at $T_c$ while preserving the single-particle gap, thereby opening a robust pseudogap regime. Our simulations quantitatively reproduce the key spectroscopic features observed in the metallic $d$-wave altermagnet KV$_2$Se$_2$O, revealing a unique microscopic picture that connects altermagnetic Fermi-surface reconstruction, unconventional density-wave order, and pseudogap phenomena.

{\sl Model.---}The DFT band structure and the corresponding Fermi surfaces of KV$_2$Se$_2$O are shown in Fig.~\ref{figyc1}. The strong $d$-wave altermagnetic spin splitting of the electron bands produces two metallic spin-selective quasi-1D Fermi-surface sheets oriented along orthogonal directions, together with additional spin-polarized hole pockets around the $X$ and $Y$ points. As evident from Fig.~\ref{figyc1}(b), the quasi-1D sheets exhibit nearly perfect nesting at ${\bf Q}=(\pi,\pi)$, which drives the dominant density-wave instability~\cite{gruner1985charge,gruner1988dynamics,lee1974conductivity,peierls1996quantum}. The closed nearly circular pockets near the $X$ and $Y$ points lack the geometric phase-space advantage associated with  nesting and  are not expected to develop a comparably strong density-wave instability, contributing mainly to the electronic background. We start with an interacting Hamiltonian~\cite{PhysRevB.97.224423}, focusing on the electron bands, 
\begin{equation}
\label{eq:gen_Hamil}
H=\sum_{{\bf k},s}\xi_{{\bf k},s}\,c^{\dagger}_{{\bf k}s}c_{{\bf k}s}
+\frac{1}{2}\sum_{{\bf k k' q}, s s'}
U_{ss'}({\bf q})\,
c^{\dagger}_{{\bf k+q}s}
c_{{\bf k}s}
c^{\dagger}_{{\bf k'}s'}
c_{{\bf k'+q}s'},
\end{equation}
where $c^{\dagger}_{{\bf k}s}$ and $c_{{\bf k}s}$ denote creation and annihilation operators for an electron with momentum ${\bf k}$ and spin $s$. The tight-binding dispersion is given by~\cite{PhysRevX.12.011028} 
$\xi_{{\bf k},s}
=-2t(\cos k_x+\cos k_y)
-4t'\cos k_x \cos k_y-2t_js_z(\cos k_x-\cos k_y)-\mu$,
where $t$ and $t'$ are the nearest- and next-nearest-neighbor hopping amplitudes, and $t_j$ parametrizes the $d$-wave momentum-dependent spin splitting induced by altermagnetism, while $\mu$ is the chemical potential.  The interaction $U_{ss'}({\bf q})$ describes the spin-conserving electron-electron interactions.   

Since the quasi-1D Fermi-surface sheets exhibit nearly perfect nesting at ${\bf Q}=(\pi,\pi)$ for both spin sectors, consistent with the dominant ordering wave vector observed experimentally, we single out the strongly enhanced density-wave interaction channel at ${\bf q}={\bf Q}$, and then,  the mean-field decoupling in the density-wave channel yields spin-resolved order parameters,
\begin{equation}
\Delta_{s}=\frac{V}{2}\sum_{\bf k}\langle{c^{\dagger}_{{\bf k+Q}s}c_{{\bf k}s}}\rangle,\quad \Delta_{s}^*=\frac{V}{2}\sum_{\bf k}\langle{c^{\dagger}_{{\bf k}s}c_{{\bf k+Q}s}}\rangle,  
\end{equation}
and the resulting mean-field Hamiltonian can be written as
\begin{eqnarray}
H_{\rm MF}&=&\sum_{s}\frac{1}{2}\sum_{\bf k}(c^{\dagger}_{{\bf k}s},c^{\dagger}_{{\bf k+Q}s})\left(\begin{array}{cc}
\xi_{{\bf k},s} & \Delta_{s} \\
\Delta_{s}^*& \xi_{{\bf k+Q},s} 
\end{array}\right)\left(\begin{array}{c}
c_{{\bf k}s}\\
c_{{\bf k+Q}s}
\end{array}\right)\nonumber\\
&&\mbox{}-|\Delta_{\uparrow}|^2/V-|\Delta_{\downarrow}|^2/V,
\end{eqnarray}
where $V=4U_{ss}({\bf Q})$ denotes the dominant intra-spin density-wave interaction, and the inter-spin interaction $U_{s\bar{s}}({\bf Q})$ is much weaker due to the small orbital/sublattice overlap between the two spin-selective quasi-1D Fermi-surface sheets,  $U_{s\bar{s}}({\bf Q})\ll{U_{ss}({\bf Q})}$ (see Sec.~SII~B), and therefore provides only a subleading correction to the density-wave instability. Diagonalizing the electronic sector in the mean-field Hamiltonian $H_{\rm MF}$, we obtain the spin-resolved quasiparticle spectrum,
\begin{equation}\label{QE}
E_{{\bf k}s}^{\pm}
=
\frac{\xi_{{\bf k},s}+\xi_{{\bf k}+{\bf Q},s}}{2}
\pm
\sqrt{
\left(
\frac{\xi_{{\bf k},s}-\xi_{{\bf k}+{\bf Q},s}}{2}
\right)^2
+
|\Delta_{s}|^2
},
\end{equation}
 and the self-consistent gap equation takes the form (Sec.~SII):
\begin{equation}\label{GE}
\frac{1}{V}
=
\frac{1}{2}
\sum_{\bf k}
\frac{
f\!\left(E_{{\bf k}s}^{-}\right)
-
f\!\left(E_{{\bf k}s}^{+}\right)}{2\sqrt{[(\xi_{{\bf k},s}-\xi_{{\bf k}+{\bf Q},s})/{2}]^2
+|\Delta_{s}|^2
}}, 
\end{equation}
where $f(x)$ is the Fermi-Dirac distribution function.

By solving self-consistently Eqs.~(\ref{QE}) and (\ref{GE}), one can determine the spin-resolved density-wave gaps $|\Delta_s|$. In realistic calculation, a simple fit to the DFT band structure yields the parameters $t=90.375$ meV, $t'=-36.9375$ meV, $t_j=194.24$ meV, and $\mu=75.25$ meV. The resulting tight-binding result reproduces the DFT band structure, as shown in Fig.~\ref{figyc1}(b). The interaction strength $V$ is fixed to reproduce the experimentally observed low-temperature gap magnitude, $|\Delta_s(T=20~\mathrm{K})|=35~\mathrm{meV}$~\cite{Jiang2025}. Once $V$ is fixed, all remaining physical properties become genuine outputs of the self-consistent calculation within the DFT-derived model. 

\begin{figure}[htb]
  \includegraphics[width=8.6cm]{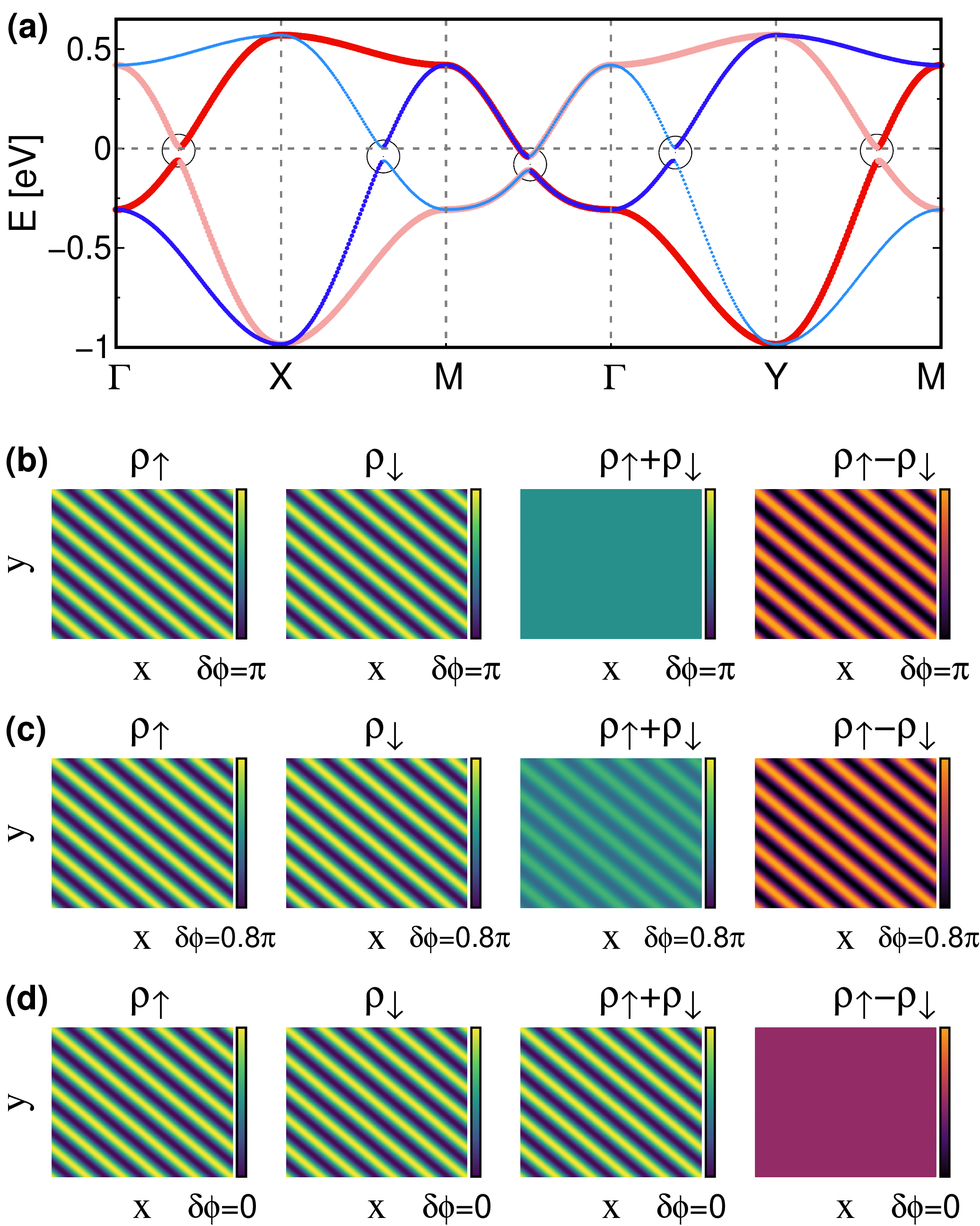}
  \caption{
  Mean-field results based on the tight-binding model at $T=0$.
  ({\bf a}) Quasiparticle energy spectra in the density-wave state.
  Red (blue) curves denote the spin-up (spin-down) bands. The darker thick
  curves represent the original bands, whereas the lighter thin curves
  correspond to the backfolded bands. Pronounced gap openings of approximately
  $35$~meV develop at the intersections between the original and backfolded
  bands, as highlighted by the black dashed circles.
  ({\bf b})--({\bf d}) Real-space modulations of the spin-resolved electronic
  densities $\rho_{\uparrow}({\bf r})$ and $\rho_{\downarrow}({\bf r})$,  the total charge-density modulation
  $\rho_{\uparrow}({\bf r})+\rho_{\downarrow}({\bf r})$ and spin-density
  modulation $\rho_{\uparrow}({\bf r})-\rho_{\downarrow}({\bf r})$, for
  relative phases $\delta\phi=\phi_{\uparrow}-\phi_{\downarrow}=\pi$, $0.8\pi$,
  and $0$, respectively. These configurations correspond to a pure stripe
  CDW, an intertwined stripe charge and spin density wave,
  and a pure stripe SDW, respectively.
  }
  \label{figyc2}
\end{figure}

{\sl Ground State.---}The numerical mean-field results for the $T=0$ ground state are shown in
Fig.~\ref{figyc2}. As seen from the quasiparticle spectrum in
Fig.~\ref{figyc2}(a) [Eq.~(\ref{QE})], upon the formation of the
density-wave order, pronounced gap openings emerge near the Fermi level. Specifically, in the absence of gap opening, the spectral-weight bands in Fig.~\ref{figyc2}(a) (the dark, thick curves) reduce directly to the original DFT bands shown in Fig.~\ref{figyc1}(c), whereas the weaker spectral branches (lighter, thinner curves) arise from band folding
associated with the ordering vector ${\bf Q}=(\pi,\pi)$.  For example, in
the spin-down sector (blue), these weaker branches originate from the
corresponding DFT bands translated by ${\bf Q}$: the branches along the
$\Gamma$--$X$, $X$--$M$, $M$--$\Gamma$, and $\Gamma$--$Y$ segments in
Fig.~\ref{figyc2}(a) originate, respectively, from the original DFT bands
[Fig.~\ref{figyc1}(c), or equivalently the dark, thick curves in
Fig.~\ref{figyc2}(a)] along the $M$--$Y$, $Y$--$\Gamma$,
$\Gamma$--$M$, and $M$--$X$ segments. The gap opening occurs precisely at the crossing points between the original and backfolded bands. Consequently, the spin-up sector exhibits a gap along $\Gamma$-X direction and the spin-down sector shows gap openings along $\Gamma$-Y and X-M, while a spin-degenerate gap opening is observed along $\Gamma$-M. This characteristic spin- and momentum-selective gap structure
emerges without introducing any additional explicit symmetry breaking and
is in good agreement with the ARPES measurements at 20~K~\cite{Jiang2025},
which report quasiparticle gap openings and concomitant band reconstruction
along the $\Gamma$--$X/Y$ and $Y/X$--$M$ directions.

The density-wave order here gives rise to a periodic stripe modulation of the spin-resolved electronic charge density.
\begin{align}
&\rho_{s}({\bf r})
=
\rho_{s}({\bf Q})\,e^{i{\bf Q}\cdot{\bf r}}
\!+\! \text{c.c.}=
\sum_{\bf k}
\langle
c^{\dagger}_{{\bf k}+{\bf Q}s}
c_{{\bf k}s}
\rangle
\,e^{i{\bf Q}\cdot{\bf r}}
+ \text{c.c.}
\nonumber\\
&={2V^{-1}\Delta_{s}}\,e^{i{\bf Q}\cdot{\bf r}}
+ \text{c.c.}
=
{4{V}^{-1}|\Delta_{s}|}
\cos\!\left({\bf Q}\cdot{\bf r}+\phi_{s}\right).
\end{align}
Here, the complex order parameter is written as $\Delta_{s}=|\Delta_{s}|e^{i\phi_{s}}$, with $\phi_{s}$ the corresponding phase [the sliding (translational) phase].  For repulsive Coulomb interactions, the density modulations of the two spin sectors generally favor an antiphase configuration, $\phi_{\uparrow}-\phi_{\downarrow}=\pi$, which minimizes the modulation of the total charge density. As a direct consequence, the superposition of the two spin-resolved stripe orders yields a pure stripe spin density wave with vanishing charge-density modulation [as shown in Fig.~\ref{figyc2}(b)]. Conversely, the in-phase configuration, $\phi_{\uparrow}-\phi_{\downarrow}=0$, corresponds to a pure stripe charge density wave with no spin-density modulation [as shown in Fig.~\ref{figyc2}(d)]. More generally, the experimentally realized state may acquire a nontrivial
relative phase in the presence of quenched disorder, local defects,
additional interactions, or symmetry-breaking perturbations. In particular,
owing to the intrinsic locking between spin and crystallographic directions
in a $d$-wave altermagnet, structural distortions, local strain, or other
spatially anisotropic perturbations may couple differently to the two
spin-resolved condensates and shift their relative phase away from the ideal
Coulomb-selected value. For a nontrivial relative phase, charge- and
spin-density modulations coexist, giving rise to intertwined stripe charge
and spin density waves [Fig.~\ref{figyc2}(c)].   Importantly, any deviation of the relative phase from $\pi$ can generate a finite CDW component, suggesting that a stripe charge modulation may be experimentally accessible in realistic KV$_2$Se$_2$O samples through spatially resolved probes, particularly in samples with appreciable disorder or strain.

\begin{figure}[htb]
  \includegraphics[width=8.7cm]{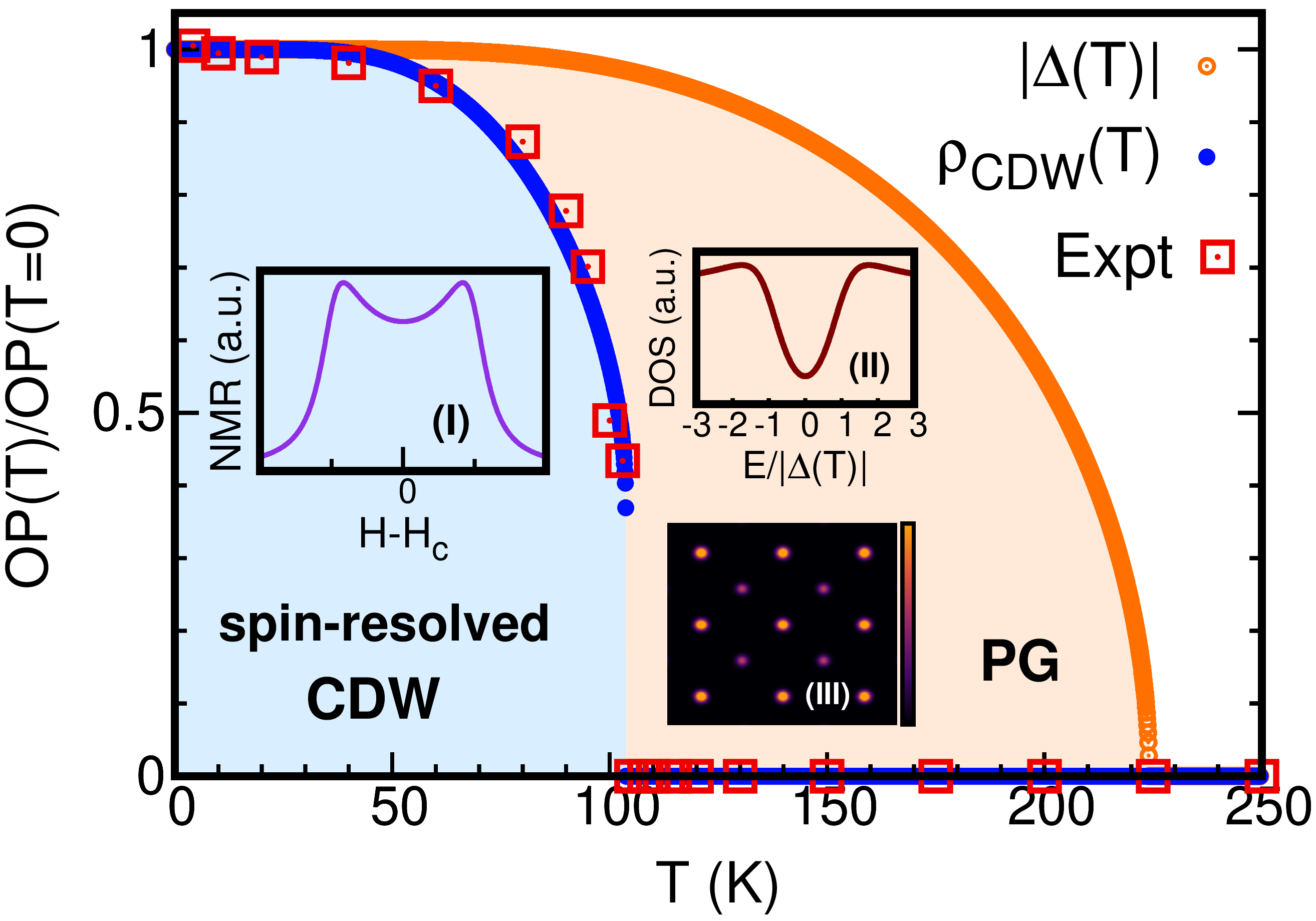}
  \caption{
  Temperature dependence of the spin-resolved density-wave modulation amplitude
  $\rho_{\rm CDW}(T)=4V^{-1}|\Delta_s(T)|
  \exp[-\langle\delta\phi_s^2(T)\rangle/2]$
  and the gap magnitude $|\Delta(T)|=|\Delta_s(T)|$.
  All physical quantities are normalized by their respective $T=0$ values,
  showing the thermal evolution of experimentally accessible observables upon
  heating and across the phase transitions.
  Experimental data (squares) are taken from Ref.~\cite{Jiang2025} and extracted
  from the temperature dependence of the spectral splitting observed in NMR
  measurements, assuming that the splitting is proportional to the
  spin-resolved density order $\rho_{\rm CDW}$.
  Inset (I): Calculated saddle-shaped NMR line profile arising from stripe
  spin-density modulation, providing a characteristic signature of the
  spin-resolved CDW at low temperature.
  Inset (II): Calculated density of states (DOS), showing the opening of the
  electronic gap in the pseudogap (PG) phase.
  Inset (III): Predicted X-ray diffraction (XRD) pattern in the PG phase.
  The bright peaks correspond to the primary lattice Bragg reflections,
  while the weaker light-purple satellite peaks originate from the CDW
  modulation.
  Specific parameters used in the calculation are presented in Sec.~SIV.
  }
  \label{figyc3}
\end{figure}

The resulting stripe spin-density modulation provides a natural microscopic origin for the characteristic NMR response. The spatially modulated local magnetic field associated with the spin density wave produces a continuous distribution of local resonance frequencies, rather than a set of discrete resonance lines, giving rise to a characteristic saddle-shaped NMR line profile, as shown in the inset (I) of Fig.~\ref{figyc3}. This behavior is consistent with the NMR measurements~\cite{Jiang2025}, where a broad and continuous saddle-shaped spectrum develops around the original central resonance upon cooling below $100$~K.

{\sl Phase transitions.---}The finite-temperature behavior and the associated phase transitions require a treatment beyond mean-field theory, as thermal phase fluctuations of the order parameter in low-dimensional electronic systems  become increasingly important at $T>0$. In systems where the phase transition is driven purely by electron-electron interactions, the amplitude of the order parameter remains well captured by the mean-field solution~\cite{yang2025efficient,PhysRevB.97.224423,PhysRevB.70.214531,PhysRevB.97.054510} whereas the phase degree of freedom must be treated by adopting a bosonized description of the collective phase mode (Nambu-Goldstone boson~\cite{nambu1960quasi,goldstone1961field,goldstone1962broken,nambu2009nobel}, i.e., phason).

We decompose the phase as $\phi_s=\phi^e_s+\delta\phi_s$ where $\delta\phi_s$ denotes phase fluctuations around equilibrium configuration $\phi^e_s$, and hence, the average of density-wave order reads (Sec.~SIII)
\begin{eqnarray}
&&\langle\rho_s({\bf r})\rangle={4V^{-1}|\Delta_s|}
\langle\cos\!\left({\bf Q}\!\cdot{\bf r}+\phi_{s}\right)\rangle\nonumber\\
&&\mbox{}~~={4V^{-1}|\Delta_s|}
\exp({-\langle{\delta\phi^2_s}\rangle}/2)\cos\!\left({\bf Q}\!\cdot{\bf r}+\phi^e_{s}\right),
\end{eqnarray}
in which a Debye-Waller-like factor~\cite{kittel1963quantum,sakurai2020modern} $\exp({-\langle{\delta\phi^2_s}\rangle}/2)$ emerges from phase fluctuations (Sec.~SIII~A). Then, the temperature dependence of the density-wave order $\rho_s({\bf r})$ and the phase transitions are governed not only by the gap, but also by thermal phase fluctuations. {This separation naturally accounts for the suppression of long-range coherence at finite temperatures, even in the presence of a robust single-particle gap. Observation of $\rho_s({\bf r})$ then requires the simultaneous presence of a finite gap amplitude $|\Delta_s|$ and long-range phase coherence.} The  average of phase fluctuations takes the form of a standard bosonic excitation~\cite{yang2025efficient,yang2025microscopic,yang2025preformed,yang2025tractable,PhysRevB.24.2751} (see Sec.~SIII~C for the detailed microscopic derivation within path-integral approach),
\begin{equation}\label{F1}
\langle\delta\phi^2_s\rangle/2=\int{\frac{d{\bf q}}{(2\pi)^2}}\frac{2n_B(\Omega_s({\bf q}))+1}{D_s\Omega_s({\bf q})},
\end{equation}
where $D_s$ is the phason inertia (temporal stiffness, see Sec.~SI),  $n_B(x)$ is the Bose-Einstein distribution, and the energy spectrum of the bosonic phason $\Omega_s({\bf q})$ is given by (see Sec.~SIII~B)
\begin{equation}\label{F2} 
\Omega^2_s(q)=m_p^2(T)+f_c(T)({\bf v}_s\cdot{\bf q})^2,
\end{equation}
where $f_c(T)$ is the condensation fraction (see Sec.~SI) and ${\bf v}_s$ is the phason velocity. {While the phason dispersion is highly anisotropic here, the phase-fluctuation field itself remains a 2D bosonic mode and all fluctuation contributions are evaluated through a 2D  momentum-space integration.} The $m_p$
represents the excitation gap of the phason, which originates from  the pinning of the density-wave order due to impurities~\cite{PhysRevB.17.535}, commensurability effect or lattice imperfections~\cite{gruner1985charge,gruner1988dynamics}. The pinning strength obeys a self-consistent temperature dependence  (see Sec.~SIII~B for detailed derivations)
\begin{equation}
m^2_p(T)=m_p^2(0)\exp({-\langle{\delta\phi^2_s(T)}\rangle}/2)|\Delta(T)|/|\Delta(0)|,
\end{equation}
which reflects the feedback between thermal phase fluctuations and the effective pinning and describes the phason softening as temperature increases. When the phason softens and becomes nearly gapless upon approaching the critical temperature, the absence of an excitation gap leads to significant phase fluctuations. Then, the density-wave order is expected to be rapidly suppressed in the vicinity of this transition. 

Solving self-consistently the gap [Eq.~(\ref{GE})] and fluctuation [Eqs.~(\ref{F1})-(\ref{F2})] equations as well as $f_c(T)$ [Eq.~(S25)] leads to the finite-temperature phase evolution and phase transitions shown in Fig.~\ref{figyc3}. As the temperature increases above 50~K, the progressively enhanced thermal phase fluctuations lead to a rapid suppression of the density-wave order $\rho_{\rm CDW}(T)$, while the gap $|\Delta(T)|$ exhibits a slow decrease. The former behavior is governed by bosonic thermal fluctuations, whereas the latter is primarily controlled by the thermal occupation of fermionic quasiparticles. Thus, the  disappearance of the density-wave ordered state at finite temperatures is controlled by the significantly enhanced thermal phase fluctuations as the phason softens toward a nearly gapless mode, rather than by a collapse of the mean-field gap amplitude. This provides a natural mechanism for a phase-disordered yet gapped state, giving rise to a pseudogap regime above the transition temperature, where the density-wave order $\rho_{\rm CDW}$ vanishes while the single-particle excitation gap remains finite and thus persists as an observable gap feature [as shown in inset (II) of Fig.~\ref{figyc3}] by spectroscopic probes such as STM and ARPES.  The density-wave order $\rho_{\rm CDW}(T)$ melts at
$T_c\approx100~$K, with a pronounced suppression in the vicinity of~$T_c$, whereas the gap opening $|\Delta(T)|$ persists, giving rise to a robust pseudogap regime, which eventually closes at a higher temperature $T_g\approx225~$K.  These features, including the temperature dependence of the normalized $\rho_{\rm CDW}(T)$ {from $T=0$ to $T_c$} as well as the values of $T_c$ and $T_g$ exhibit quantitative agreement with the experimental observations~\cite{Jiang2025}, where NMR signatures of the density-wave order vanish above 100~K, while the single-particle gap observed in ARPES is estimated to remain finite up to approximately 230-260~K in altermagnetic KV$_2$Se$_2$O.

\begin{widetext}
  \begin{center}
\begin{figure}[h]
  {\includegraphics[width=16.8cm]{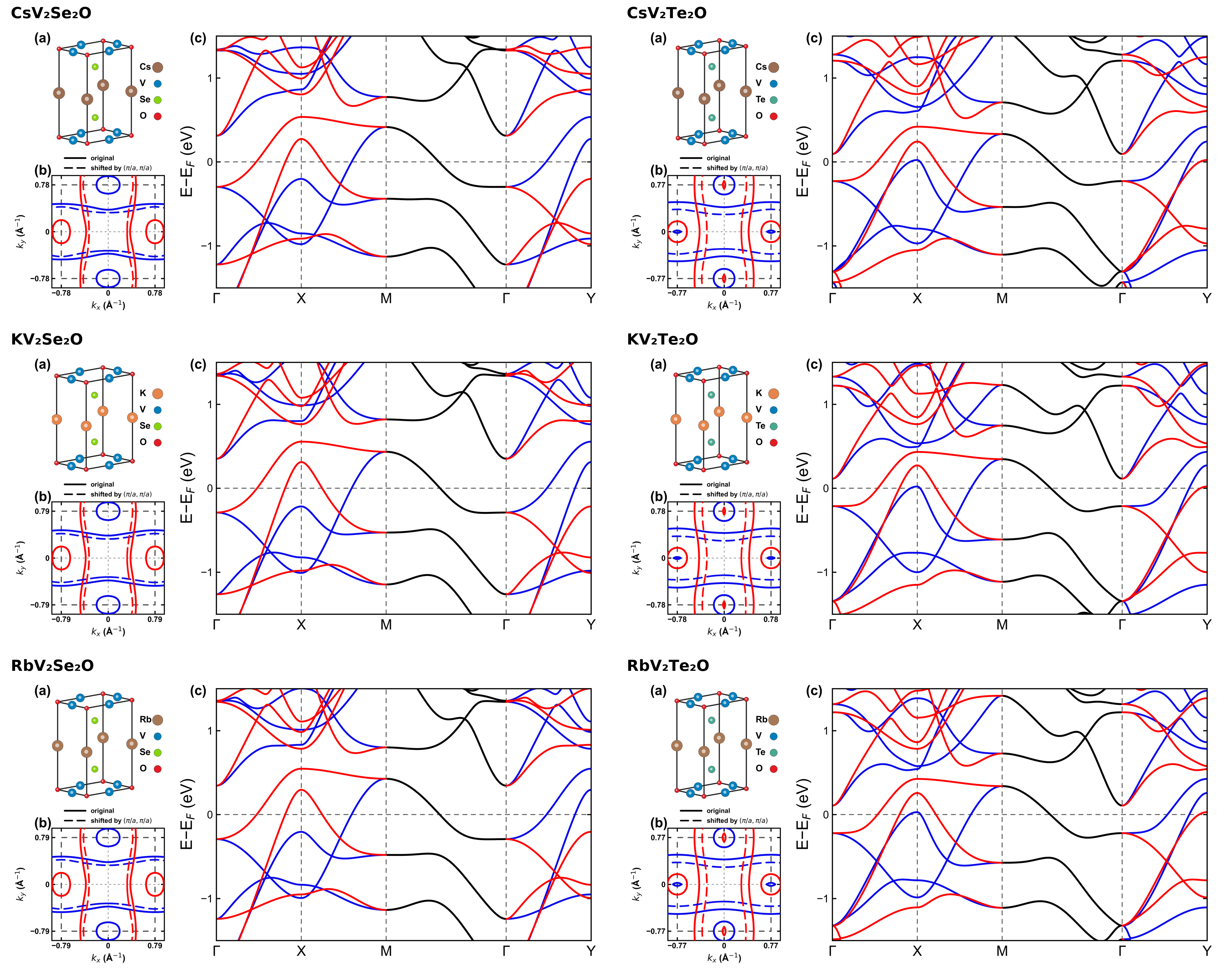}}
\caption{Atomic and electronic structures of the $AV_2X_2O$ family, with $A=\mathrm{Cs}$, $\mathrm{K}$, and $\mathrm{Rb}$ and $X=\mathrm{Se}$ and $\mathrm{Te}$, including CsV$_2$Se$_2$O, KV$_2$Se$_2$O, RbV$_2$Se$_2$O, CsV$_2$Te$_2$O, KV$_2$Te$_2$O, and RbV$_2$Te$_2$O. For each compound, ({\bf a}) crystal structure of the unit cell, ({\bf b}) Fermi surface in the $k_z=0$ plane, and ({\bf c}) band structure. Red, blue, and black curves denote spin-up, spin-down, and spin-degenerate bands, respectively. The dashed curves in ({\bf b}) represent the Fermi surfaces translated by the nesting vector ${\bf Q}=(\pi,\pi)$. The nearly parallel segments between the original and translated Fermi surfaces indicate near-perfect Fermi-surface nesting at ${\bf Q}=(\pi,\pi)$ and hence density-wave instabilities.}
\label{figyc4}
\end{figure}
\end{center}
\end{widetext}

{\sl Discussion.---}Through a self-consistent microscopic calculation, the present study reveals a spin-resolved electronic CDW instability, in which the strong $d$-wave spin-momentum locking naturally allows the two spin sectors to develop distinct CDW instabilities. This spin-resolved CDW instability has no direct analogue in conventional magnetic systems: in ferromagnets, although the electronic bands are spin split, the splitting does not exhibit the characteristic momentum-dependent sign reversal of altermagnetism, whereas conventional compensated antiferromagnets generally retain spin-degenerate electronic bands. The developed beyond-mean-field simulations here naturally reproduce several key experimental observations from ARPES and NMR measurements in a recent study of the $d$-wave altermagnet KV$_2$Se$_2$O~\cite{Jiang2025}. 

More broadly, the emergent phase degree of freedom and the pseudogap phenomenology identified here qualitatively enriches the collective dynamics of the density-wave state. An immediate experimental consequence concerns the accompanying lattice response. Since an electronic CDW generally couples to a lattice distortion at the same ordering wave vector, their lowest-order coupling can be written as
\begin{equation}
F_{\rm el-ph}=\frac{K_{\mathbf Q}}{2}|u_{\mathbf Q}|^2
+
\lambda_{\mathbf Q}
\left(
u_{\mathbf Q}^{*}\Delta(\mathbf Q)
+{\rm c.c.}
\right),
\end{equation}
where $u_{\mathbf Q}$ denotes the lattice displacement and $\Delta(\mathbf Q)$ the electronic CDW order parameter. Minimization with respect to $u_{\mathbf Q}$ immediately gives $|u_{\mathbf Q}|\propto|\Delta(\mathbf Q)|$. Therefore, once the CDW gap opens and a finite CDW amplitude develops below $T_g$, a corresponding lattice distortion should emerge simultaneously. This remains the case even though the associated electronic charge-density modulation has so far been directly observed only below $T_c$. We therefore predict that CDW-related superlattice signatures, as shown in the inset (III) of Fig.~\ref{figyc3} should become detectable by X-ray diffraction (XRD) already below $T_g$, rather than appearing only below $T_c$. The temperature onset of the XRD superlattice signal thus provides a particularly direct experimental test of whether the pseudogap originates from a preformed electronic CDW order parameter.

A second characteristic consequence arises from the qualitatively different collective phase dynamics above and below $T_c$. In the pseudogap phase, the phason remains gapless, whereas it becomes gapped upon entering the low-temperature CDW phase. The gapless phason provides an efficient low-energy scattering and decay channel for both fermionic quasiparticles and collective amplitude fluctuations, and should therefore lead to substantially stronger damping in the pseudogap regime. This difference should be directly reflected in the single-particle excitation spectrum. Although the CDW gap is already open in the pseudogap regime $T_c<T<T_g$, the strong damping is expected to smear the spectral edges and suppress sharp quasiparticle coherence peaks, producing a pseudogapped density of states with broad spectral features, as shown in inset (II) of Fig.~\ref{figyc3}. The conventional BCS-like gap-opening behavior, characterized by well-defined sharp coherence peaks, emerges only below $T_c$, where the gapping of the phason substantially suppresses the low-energy damping channels and restores coherent quasiparticle excitations. This behavior is consistent with the experimentally observed evolution of the spectral features across $T_c$~\cite{Jiang2025}. Moereover, the amplitudon, a gapped collective amplitude mode typically observable as a resonance in optical spectroscopy~\cite{pekker2015amplitude,matsunaga2013higgs,matsunaga2014light,shimano2020higgs}, is also expected to be heavily damped by the gapless phason above $T_c$~\cite{yang2025microscopic}, resulting in a broad or even unresolved feature. Below $T_c$, the opening of the phason gap suppresses this damping and gives rise to a sharp, well-defined amplitudon resonance at the characteristic amplitudon frequency.  

The evolution of the optical resonant response across $T_c$, the predicted onset of the XRD superlattice signal below $T_g$, and the emergence of sharp quasiparticle coherence peaks below $T_c$ provide complementary experimental probes for distinguishing the formation of the CDW amplitude from the subsequent evolution of its phase dynamics. These predictions offer a microscopic framework for understanding how altermagnetism can promote electronic instabilities and how strong phase fluctuations can generate pseudogap regimes in low-dimensional density-wave systems. The proposed mechanism for density-wave instabilities and the predicted experimental signatures, may also be realized and observed in the broader $AV_2X_2O$ ($A=\mathrm{K,Rb,Cs}$; $X=\mathrm{Se,Te}$) family, as suggested by Fig.~\ref{figyc4}, providing a broader materials platform for exploring altermagnetism-driven density-wave order and fluctuation-induced pseudogap behavior.

{\it Acknowledgments.---}This work is supported by the US Department of Energy, Office of Science, Basic Energy Sciences, under Award Number DE-SC0020145 as part of Computational Materials Sciences Program. F.Y. and L.Q.C. also appreciate the  generous support from the Donald W. Hamer Foundation through a Hamer Professorship at Penn State.

\bibliography{ref.bib}

\end{document}


\title{Emergent spin-resolved electronic density waves from strong $d$-wave altermagnetism and pseudogap phenomena from phason fluctuations  (Supplementary Materials)}

\author{Fei Yang}
\email{fzy5099@psu.edu}

\affiliation{Department of Materials Science and Engineering and Materials Research Institute, The Pennsylvania State University, University Park, PA 16802, USA}

\author{Guo-Dong Zhao}
\affiliation{Department of Materials Science and Engineering and Materials Research Institute, The Pennsylvania State University, University Park, PA 16802, USA}

\author{Binghai Yan}
\affiliation{Department of Physics, The Pennsylvania State University, University Park, PA 16802, USA}
\affiliation{The Center for Theory of Emergent Quantum Matter, The Pennsylvania State University, University Park, PA 16802, USA}

\author{Long-Qing Chen}
\email{lqc3@psu.edu}

\affiliation{Department of Materials Science and Engineering and Materials Research Institute, The Pennsylvania State University, University Park, PA 16802, USA}

\maketitle 
\tableofcontents
\section{Derivation of the effective action}

In this section, we present a microscopic derivation of the low-energy effective action starting from a fermionic Hamiltonian with the effective spin-conserving electron–electron  interactions (extended Hubbard interactions). By integrating out the fermionic degrees of freedom, we obtain a controlled description of the density-wave order parameter and its associated collective fluctuations in the ordered state.  For notational simplicity, throughout the derivation we focus on a single spin sector and suppress the explicit spin index. This choice does not entail any loss of generality, since different spin components are decoupled in the context of the spin-resolved CDWs considered here. 

Based on the Hamiltonian presented in the main text, the action of the model is given by
\begin{equation}
  S=\int{dx}\,\psi^{\dagger}(x)(i\partial_t-\xi_{\hat {p}})\psi(x)-\int{dxdx'}\frac{1}{2}U(x-x')\psi^{\dagger}(x)\psi(x)\psi^{\dagger}(x')\psi(x'),
\end{equation}
where $\psi(x)$ is the electron field operator, $\hat{\mathbf p}=-i\hbar\bm{\nabla}$ is the momentum operator, and $x=(t,\mathbf{x})$ is the space–time four-vector.

We consider an electronic density-wave order condensing at wave vector $\mathbf Q$, for which the expectation value of the electronic density takes the form~\cite{PhysRevB.97.224423}
\begin{equation}
\langle\psi^{\dagger}({x})\psi(x)\rangle=\rho_{Q}(R)e^{i{Q}\cdot{x}}+\rho_{-Q}(R)e^{-i{Q}\cdot{x}}
\end{equation}
where $R=(t,{R})$ denotes a slowly varying (long-wavelength) center-of-mass space–time four-vector~\cite{yang2025microscopic,yang21theory,PhysRevB.98.094507,yang19gauge}. Reality of the density field implies $\rho_{-\mathbf Q}(R)=\rho_{\mathbf Q}^*(R)$. Applying a mean-field decoupling in the density-wave channel~\cite{mahan2013many,abrikosov2012methods}, the action becomes 
\begin{equation}
  S=\int{dx}\,\psi^{\dagger}(x)[(i\partial_t-\xi_{\hat {p}})\psi(x)-U({Q})\rho_{Q}(R)\psi^{\dagger}(x)\psi(x)e^{i{Q}\cdot{x}}-U({Q})\rho^*_{Q}(R)\psi^{\dagger}(x)\psi(x)e^{-i{Q}\cdot{x}}]+\int{dR}U({Q})|\rho_{Q}(R)|^2,
\end{equation}
Taking $U(\mathbf Q)=V/4$ and defining the order parameter as
\begin{equation}\label{drho}
\Delta(R)=\frac{V}{2}\rho_{Q}(R),\quad\quad\Delta^*(R)=\frac{V}{2}\rho_{Q}^*(R),
\end{equation}
the action can be written in Nambu space as 
\begin{equation}
  S={\int}d{R}\Big[\sum_{k}\frac{1}{2}\big(\psi^{\dagger}_{{k}}(R),\psi^{\dagger}_{{k+Q}}(R)\big)\left(\begin{array}{cc}
    i\partial_t-\xi_{{k}-i\hbar\partial_R} & -\Delta(R) \\
    -\Delta^*(R) & i\partial_t-\xi_{{k+Q}-i\hbar\partial_R}
    \end{array}\right)\left(\begin{array}{c}
\psi_{{k}}(R)\\
\psi_{{k+Q}}(R)
\end{array}\right)+\frac{|\Delta|^2}{V}\Big].\label{ss}
\end{equation}

We then treat the density-wave gap and phase on an equal footing within the standard path-integral framework~\cite{schrieffer1964theory,littlewood81gauge,yang19gauge,yang21theory}. For convenience, we set $\hbar=1$ here and hereafter. Specifically, a general complex density-wave order parameter can be written as 
\begin{equation}\label{dgap}
\Delta(R)=|\Delta|\exp[{i\phi(R)}],
\end{equation}
with $\phi(R)=\phi_e+\delta\phi(R)$. Here, $|\Delta|$ and $\phi_e$ denote the density-wave gap and equilibrium phase configuration, respectively, while $\delta\phi(R)$ represents phase fluctuations.  Moreover, for \emph{complex}  electronic order parameter ($N=2$, where
$N$ is the dimensionality of the order parameter), amplitude fluctuations are negligible, as they are exactly canceled by short-range fluctuation terms, as demonstrated in Ref.~\cite{PhysRevB.97.224423}. This cancellation is consistent with earlier theoretical results for superconductors~\cite{PhysRevB.70.214531,PhysRevB.97.054510} and other systems with complex order parameters~\cite{PhysRevB.97.224423}, in which neither amplitude fluctuations nor short-range-fluctuations corrections modify the self-consistent order-parameter microscopic equation at the thermodynamic level, and where only long-wavelength phase fluctuations play a role.

Substituting ${k}=\delta{k}-{Q}/2$, the action can be rewritten in a symmetric form as 
\begin{equation}
  S={\int}d{R}\Big[\sum_{\delta{k}}\frac{1}{2}\big(\psi^{\dagger}_{{ \delta{k}-Q/2}}(R),\psi^{\dagger}_{{ \delta{k}+Q/2}}(R)\big)\left(\begin{array}{cc}
    i\partial_t-\xi_{{ \delta{k}-Q/2}-i\hbar\partial_{R}} & -\Delta(R) \\
    -\Delta^*(R) & i\partial_t-\xi_{{ \delta{k}+Q/2}-i\hbar\partial_{R}}
    \end{array}\right)\left(\begin{array}{c}
\psi_{{ \delta{k}-Q/2}}(R)\\
\psi_{{ \delta{k}+Q/2}}(R)
\end{array}\right)+\frac{|\Delta|^2}{V}\Big].
\end{equation}
We then apply the chiral (gauge) transformations~\cite{PhysRevB.98.094507,nambu2009nobel}
\begin{equation}
{\psi}_{ \delta{k}-Q/2}\rightarrow{\psi}_{\delta{k}-Q/2}\exp[{i\phi(R)}/2],~~~~~\psi_{\delta{k}+Q/2}\rightarrow\psi_{\delta{k}+Q/2}\exp[-{i\phi(R)}/2],
\end{equation}
and neglect trivial total-derivative terms. The action becomes
\begin{eqnarray}\label{sss}
  S&=&\int{dR}\Big\{\frac{1}{2}\sum_{\delta{k}}\Psi_{\delta{k}}^{\dagger}(R)\Big[i\partial_t-\delta\xi_{\delta{k}}-{\bar \xi}_{{\delta}k}\tau_3-|\Delta|\tau_1-\frac{\partial_t\delta\phi}{2}\tau_3+\frac{{v}_{Q/2}\cdot\partial_{R}\delta\phi}{2}-\frac{(\partial_{R}\phi\cdot{{m}^{-1}_{Q/2}}\cdot\partial_{R}\phi)}{8}\Big]\Psi_{\delta{k}}(R)+\frac{|\Delta|^2}{V}\Big] \nonumber \\
  &=&\int{dR}\Big[\frac{1}{2}\sum_{\delta{k}}\Psi_{\delta{k}}^{\dagger}(R)(G_0^{-1}-\Sigma)\Psi_{\delta{k}}(R)+\frac{|\Delta|^2}{V}\Big],~~~ 
\end{eqnarray}
where $\tau_i$ denotes the Pauli matrices in Nambu space and Nambu spinor $\Psi_{\delta{k}}^{\dagger}(R)=\big(\psi^{\dagger}_{{ \delta{k}-Q/2}}(R),\psi^{\dagger}_{{ \delta{k}+Q/2}}(R)\big)$; the Green function is determined by 
\begin{equation}
G^{-1}_0=i\partial_t-\delta\xi_{\delta{k}}-{\bar \xi}_{\delta{k}}\tau_3-|\Delta|\tau_1,
\end{equation}
with 
\begin{equation}
\delta\xi_{\delta{k}}=(\xi_{{ \delta{k}-Q/2}}+\xi_{{ \delta{k}+Q/2}})/2,\quad\quad {\bar \xi}_{\delta{k}}=(\xi_{{ \delta{k}-Q/2}}-\xi_{{ \delta{k}+Q/2}})/2,
\end{equation}
and the self-energy reads 
\begin{equation}
\Sigma={\partial_t\delta\phi}/{2}\tau_3-{{v}_{Q/2}\cdot\partial_{R}\delta\phi}/{2}+(\partial_{R}\phi/2\cdot{{m}^{-1}_{Q/2}}\cdot\partial_{R}\phi/2)/2,
\end{equation}
Here, ${v}_{Q/2}=\partial_{Q/2}\xi_{Q/2}$ the group velocity evaluated at momentum, and ${m}_{Q/2}^{-1}=\partial_{Q/2}\partial_{Q/2}\xi_{Q/2}$ represents the inverse effective-mass tensor. Then, through the standard integration over the Fermi field in the Matsubara representation~\cite{schrieffer1964theory,mahan2013many,abrikosov2012methods}, one has
\begin{equation}
  S=\int{dR}\Big\{-{\rm {\bar T}r}\ln{G^{-1}_0}+\sum_n^{\infty}\frac{1}{n}{\rm {\bar T}r}[(\Sigma{G_0})^n]+\frac{|\Delta|^2}{V}\Big\},
\end{equation}
where the Green function in the Matsubara representation: 
\begin{equation}\label{GreenfunctionMr}
G_0(ip_n,\delta{k})=\frac{ip_n\tau_0-\delta\xi_{\delta{k}}\tau_0+{\bar \xi}_{\delta{k}}\tau_3+|\Delta|\tau_1}{(ip_n-E_{\delta{k}}^+)(ip_n-E_{\delta{k}}^-)},  
\end{equation}  
with 
\begin{equation}
E_{\delta{k}}^{\pm}=\delta\xi_{\delta{k}}\pm{E_{{\delta}k}},\quad\quad E_{{\delta}k}=\sqrt{{\bar \xi}_{\delta{k}}^2+|\Delta|^2}.
\end{equation}

Further keeping the lowest two orders (i.e., $n=1$ and $n=2$) and neglecting the trivial terms lead to the result: 
\begin{equation}
  S_{\rm eff}=\int{dR}\Bigg[-\frac{1}{2}\sum_{ip_n,\delta{k}}\ln[(ip_n-E_k^+)(ip_n-E_k^-)]+\frac{|\Delta|^2}{V}+\chi_{33}\Big(\frac{\partial_t\delta\phi}{2}\Big)^2+\chi_{00}\Big(\frac{{v}_{Q/2}\cdot\partial_{R}\delta\phi}{2}\Big)^2+\chi_0\frac{(\partial_{R}\phi\cdot{{m}^{-1}_{Q/2}}\cdot\partial_{R}\phi)}{8}\Bigg]
\end{equation}
Here, the correlation coefficients related to the phase sector are given by
\begin{eqnarray}
  \chi_0&=&\frac{1}{2}\sum_{ip_n,\delta{k}}{\rm Tr}[G_0(ip_n,\delta{k})]=\frac{1}{2}\sum_{\delta{k}}[f(E_{{\delta}k}^+)+f(E_{{\delta}k}^-)], \label{chi0}\\
  \chi_{33}&=&\frac{1}{4}\sum_{ip_n,\delta{k}}{\rm Tr}[G_0(ip_n,\delta{k})\tau_3G_0(ip_n,\delta{k})\tau_3]=\frac{1}{2}\sum_{ip_n,\delta{k}}\frac{(ip_n-\delta\xi_{\delta{k}})^2+\bar\xi_{\delta{k}}^2-|\Delta|^2}{[(ip_n-\delta\xi_{\delta{k}})^2-\bar\xi_{\delta{k}}^2-|\Delta|^2]^2}\nonumber\\
  &=&\frac{1}{2}\sum_{ip_n,\delta{k}}\partial_{\bar\xi_{\delta{k}}}\Big[\frac{{\bar \xi}_{\delta{k}}}{(ip_n-E_{\delta{k}}^+)(ip_n-E_{\delta{k}}^-)}\Big]=\frac{1}{2}\sum_{\delta{k}}\partial_{\bar\xi_{\delta{k}}}\Big[{\bar \xi}_{\delta{k}}\frac{f(E_{\delta{k}}^+)-f(E_{\delta{k}}^-)}{2E_{{\delta}k}}\Big]=-D,\\
   \chi_{00}&=&\frac{1}{4}\sum_{ip_n,\delta{k}}{\rm Tr}[G_0(ip_n,\delta{k})G_0(ip_n,\delta{k})]=\frac{1}{2}\sum_{ip_n,\delta{k}}\frac{(i-\delta\xi_{\delta{k}})^2+\bar\xi_{\delta{k}}^2+|\Delta|^2}{[(ip_n-\delta\xi_{\delta{k}})^2-\bar\xi_{\delta{k}}^2-|\Delta|^2]^2}\nonumber\\
   &=&\frac{1}{2}\sum_{\delta{k}}\partial_{E_{{\delta}k}}\Big[\frac{f(E_{\delta{k}}^+)-f(E_{\delta{k}}^-)}{2}\Big]=\frac{1}{2}\sum_{\delta{k}}\frac{|\Delta|^2}{E_{\delta{k}}}\partial_{E_{{\delta}k}}\Big[\frac{f(E_{\delta{k}}^+)-f(E_{\delta{k}}^-)}{2E_{\delta{k}}}\Big]+\frac{1}{2}\sum_{\delta{k}}\partial_{\bar\xi_{\delta{k}}}\Big[{\bar \xi}_{\delta{k}}\frac{f(E_{\delta{k}}^+)-f(E_{\delta{k}}^-)}{2E_{{\delta}k}}\Big].~~~~~
\end{eqnarray}  
Mathematically, by decomposing $\delta{k}=\delta k_\parallel \hat{Q}+\delta{k}_\perp$ with
$\delta k_\parallel\equiv \delta{k}\cdot \hat{Q}$, and linearizing the nested bands within a low-energy patch,
$\bar\xi_{\delta{k}}\simeq v_\parallel\,\delta k_\parallel$ with $v_\parallel\equiv {v}_{{Q}/2}\!\cdot\!\hat{Q}$,
one has $\partial_{\bar\xi_{\delta{k}}}=(1/v_\parallel)\partial_{\delta k_\parallel}$.
Then one obtains
\begin{align}
&{v}_{Q/2}{v}_{Q/2}\frac{1}{2}\sum_{\delta{k}}\partial_{\bar\xi_{\delta{k}}}
\Big[{\bar \xi}_{\delta{k}}\frac{f(E_{\delta{k}}^+)-f(E_{\delta{k}}^-)}{2E_{\delta{k}}}\Big]
+\frac{{m}^{-1}_{Q/2}}{2}\frac{1}{2}\sum_{\delta{k}}
\big[f(E_{\delta{k}}^+)+f(E_{\delta{k}}^-)\big]\nonumber\\
&=\frac{1}{2}\!\int\! d\delta k_\perp
\!\int_{-\Lambda_\parallel}^{\Lambda_\parallel}\! d\delta k_\parallel\,
\frac{{v}_{Q/2}{v}_{Q/2}}{v_\parallel}\partial_{\delta k_\parallel}
\Big[{\bar \xi}_{\delta{k}}\frac{f(E_{\delta{k}}^+)-f(E_{\delta{k}}^-)}{2E_{\delta{k}}}\Big]
+\frac{1}{2}\!\int\! d\delta k_\perp
\!\int_{-\Lambda_\parallel}^{\Lambda_\parallel}\! d\delta k_\parallel\,
\frac{{m}^{-1}_{Q/2}}{2}
\frac{f(E_{\delta{k}}^+)+f(E_{\delta{k}}^-)}{2}\nonumber\\
&=\frac{1}{2}\!\int\! d\delta k_\perp
\Bigg\{\frac{{v}_{Q/2}{v}_{Q/2}}{v_\parallel}
\Big[{\bar \xi}_{\delta{k}}\frac{f(E_{\delta{k}}^+)-f(E_{\delta{k}}^-)}{2E_{\delta{k}}}
\Big]_{-\Lambda_\parallel}^{+\Lambda_\parallel}
+\int_{-\Lambda_\parallel}^{\Lambda_\parallel}\! d\delta k_\parallel\,
\frac{{m}^{-1}_{Q/2}}{2}
\frac{f(E_{\delta{k}}^+)+f(E_{\delta{k}}^-)}{2}
\Bigg\}.
\end{align}
The above combination is constrained by the Ward identity associated with translational invariance
(equivalently, charge conservation in the chiral basis)~\cite{peskin2018introduction}, i.e., the density-wave analog of the paramagnetic-diamagnetic cancellation
in the longitudinal response~\cite{schrieffer1964theory}. After subtracting the normal-state contribution, the boundary term vanishes at
$|\bar\xi|\sim \Lambda_E\equiv v_\parallel\Lambda_\parallel$ and exactly cancels the diamagnetic piece in the perfectly nested and linearized limit.
Away from this limit, the residual is controlled by imperfect nesting, and is parametrically small,
scaling as $\mathcal{O}(|\Delta|^2/\Lambda_E^2)\sim\mathcal{O}(|\Delta|^2/E_F^2)$.\\

Consequently, the final result is written as
\begin{equation}\label{ace}
S_{\rm eff}=S_{\rm gap}+S_{\rm phase},
\end{equation}
with the gap-related part 
\begin{equation}
S_{\rm gap}=\int{dR}\Bigg[-\frac{1}{2}\sum_{ip_n,\delta{k}}\ln[(ip_n-E_k^+)(ip_n-E_k^-)]+\frac{|\Delta|^2}{V}\Bigg],
\end{equation}
and the phase-related one
\begin{equation}
S_{\rm phase}=\int{dR}\Bigg[-D\Big(\frac{\partial_t\delta\phi}{2}\Big)^2+Df_c\Big(\frac{{v}_{Q/2}\cdot\partial_{R}\delta\phi}{2}\Big)^2\Bigg],
\end{equation}
where temporal stiffness $D$ is given by
\begin{equation}\label{DEq}
D=-\frac{1}{2}\sum_{\delta{k}}\partial_{\bar\xi_{\delta{k}}}\Big[{\bar \xi}_{\delta{k}}\frac{f(E_{\delta{k}}^+)-f(E_{\delta{k}}^-)}{2E_{{\delta}k}}\Big],
\end{equation}
and the condensation fraction $f_c$~\cite{hayashi1996topological,hayashi2000ginzburg,yang2025microscopic} is determined by 
\begin{equation}\label{fcEq}
f_c=\frac{1}{2D}\sum_{\delta{k}}\frac{|\Delta|^2}{E_{\delta{k}}}\partial_{E_{{\delta}k}}\Big[\frac{f(E_{\delta{k}}^+)-f(E_{\delta{k}}^-)}{2E_{\delta{k}}}\Big].
\end{equation}

\section{Gap opening}

\subsection{Derivation of gap equation}

In this part, we derive the self-consistent density-wave gap equation. Starting from Eq.~(\ref{ace}), we impose the equilibrium condition by minimizing the effective action with respect to the density-wave order parameter, i.e., $\partial_{|\Delta|}S_{\rm gap}=0$. This yields 
\begin{eqnarray}\label{SMGE}
\frac{1}{2}\sum_{ip_n,\delta{k}}\frac{2|\Delta|}{(ip_n\!-\!E_k^+)(ip_n\!-\!E_k^-)}+\frac{2|\Delta|}{V}\!=\!0 \quad\Rightarrow\quad \frac{1}{V}=-\frac{1}{2}\sum_{ip_n,\delta{k}}\frac{1}{(ip_n\!-\!E_k^+)(ip_n\!-\!E_k^-)}=\frac{1}{2}\sum_{\delta{k}}\frac{f(E_{\delta{k}}^-)-f(E_{\delta{k}}^+)}{2E_{{\delta}k}},
\end{eqnarray}  
where $f(x)$ is the Fermi–Dirac distribution function and the quasiparticle excitation spectrum is written as 
\begin{equation}
E_{\delta{k}}^{\pm}=(\xi_{{ \delta{k}-Q/2}}+\xi_{{ \delta{k}+Q/2}})/2\pm{E_{{\delta}k}},\quad\quad E_{{\delta}k}=\sqrt{[(\xi_{{ \delta{k}-Q/2}}-\xi_{{ \delta{k}+Q/2}})/2]^2+|\Delta|^2}.
\end{equation}
By shifting the momentum variable according to  ${k}=\delta{k}-{Q}/2$, the gap equation presented in the main text is recovered.

We note that in deriving the self-consistent gap equation above, phase fluctuations are not explicitly incorporated into the fermionic spectrum, i.e., the Doppler-shift effect  ($\partial_{|\Delta|}S_{\rm phase}$) associated with phase-induced momentum shifts~\cite{yang2025efficient,yang2025preformed,yang2025tractable,PhysRevB.98.094507,yang21theory} is neglected. This approximation should be regarded as a controlled heuristic approximation for electronic-density-wave systems with a large phase-mode velocity, as commonly adopted in the existing literature. This is because with a large phase-mode velocity,  the phase space available for the thermal average of Doppler-shift effects $\langle(\partial_{R}\delta\phi/{2})^2\rangle\sim\langle(q^2\delta\phi_{q}/{2})^2\rangle$ is strongly restricted. As a result, the corresponding corrections from thermal phase fluctuations to the gap equation are parametrically small and do not qualitatively modify the mean-field solution at finite temperature~\cite{yang2025efficient,yang2025preformed,yang2025tractable,PhysRevB.98.094507,yang21theory}.  The explicit inclusion of this effect
 is not the focus of the present study and will be discussed in a future publication.  By contrast, such Doppler-shift effects may become non-negligible in lattice-driven density-wave systems, where the phason energy spectrum is strongly renormalized by the ratio $m/m^*$~\cite{Lee1974conductivity,gruner1988dynamics,gruner1985charge}. This renormalization leads to a substantial suppression of the phason mode velocity, thereby significantly enlarging the phase space available for the thermal averaging of Doppler-shift effects. In this regime, phase fluctuations can strongly feed back into the fermionic sector and modify the self-consistent gap equation beyond the mean-field level~\cite{yang2025microscopic}.

\subsection{Microscopic origin of the spin-dependent effective interaction}
\label{sec:microscopic_interaction}

In the main text, the electron-electron interaction projected onto the
low-energy altermagnetic bands is described by $U_{ss'}({\bf q})$.
Here we discuss its microscopic origin and show explicitly how the
real-space screened Coulomb interaction, orbital structure, and
$d$-wave altermagnetism of KV$_2$Se$_2$O enter the effective
density-wave interaction.

Specifically, the low-energy electronic states of KV$_2$Se$_2$O are
predominantly derived from the V-$3d$ orbitals. Of particular importance
for the density-wave instability are the strongly anisotropic
$d_{xz}$- and $d_{yz}$-dominated bands. Because of their directional
orbital characters, the corresponding electronic states naturally
possess strongly anisotropic dispersions along two orthogonal
crystallographic directions. In the altermagnetic state, the orbital
and sublattice characters are strongly correlated with the spin
polarization. We denote the two dominant low-energy orbital-sublattice
sectors by
\begin{equation}
|X\rangle \equiv |A,d_{xz}\rangle,
\qquad
|Y\rangle \equiv |B,d_{yz}\rangle,
\label{eq:S_XY_definition}
\end{equation}
where $A$ and $B$ denote the two symmetry-related magnetic sublattice
sectors.

The dominant components of the two spin-resolved quasi-one-dimensional
bands can then be schematically represented as
\begin{align}
|u_{\uparrow}({\bf k})\rangle
=
\sqrt{1-\eta_{\bf k}^{2}}\,|X\rangle
+
\eta_{\bf k}\,|Y\rangle
+\cdots,\quad
|u_{\downarrow}({\bf k})\rangle
=
\sqrt{1-\eta_{\bf k}^{2}}\,|Y\rangle
+
\eta_{\bf k}\,|X\rangle
+\cdots ,
\label{eq:S_wavefunctions}
\end{align}
where $\eta_{\bf k}\ll1$ characterizes the minority
orbital-sublattice component on the nested portions of the Fermi
surface. The ellipses represent additional orbital components carrying
only small spectral weight in the relevant low-energy bands.

Equation~(\ref{eq:S_wavefunctions}) represents the strong
spin-orbital-sublattice locking of the altermagnetic state: the
spin-up nested band predominantly occupies the $X$ sector, whereas
the spin-down nested band predominantly occupies the symmetry-related
$Y$ sector. In the ideal locking limit $\eta_{\bf k}\rightarrow0$,
the two sectors become completely spin selective. The two dominant
states are related by the crystalline operation associated with the
$d$-wave altermagnetic symmetry,
\begin{equation}
|X,\uparrow\rangle
\longleftrightarrow
|Y,\downarrow\rangle.
\label{eq:S_AM_relation}
\end{equation}

The $d_{xz}$-dominated sector preferentially disperses along one
crystallographic direction, whereas the $d_{yz}$-dominated sector
preferentially disperses along the orthogonal direction. The strong
$d$-wave altermagnetic splitting further enhances this anisotropy,
leading to the spin-selective quasi-one-dimensional Fermi surfaces
shown in the main text. Despite their different spin-orbital
characters, both Fermi-surface sheets exhibit strong nesting at the
same wave vector ${\bf Q}=(\pi,\pi)$. Consequently, both sectors
possess an enhanced particle-hole susceptibility at ${\bf Q}$ and
are susceptible to density-wave formation.

We now derive the effective interaction starting from the screened
electron-electron interaction in real space. The interaction Hamiltonian
can be written as
\begin{equation}
H_C
=
\frac{1}{2}
\sum_{ss'}
\int d{\bf r}\,d{\bf r}'\,
\psi_s^\dagger({\bf r})
\psi_{s'}^\dagger({\bf r}')
W({\bf r},{\bf r}')
\psi_{s'}({\bf r}')
\psi_s({\bf r}),
\label{eq:S_realspace_Coulomb}
\end{equation}
where $W({\bf r},{\bf r}')$ denotes the screened Coulomb interaction.
For a translationally invariant system, it depends only on the relative
coordinate, 
$W({\bf r},{\bf r}')
=
W({\bf r}-{\bf r}')
=
\int \frac{d^dq}{(2\pi)^d}
W({\bf q})
e^{i{\bf q}\cdot({\bf r}-{\bf r}')}$.  To project Eq.~(\ref{eq:S_realspace_Coulomb}) onto the relevant
low-energy orbital-sublattice sectors, we expand the electron field
operator in localized Wannier orbitals,
\begin{equation}
\psi_s({\bf r})
=
\sum_{{\bf R},a}
w_a({\bf r}-{\bf R}-{\boldsymbol\tau}_a)
d_{{\bf R}as},
\qquad
a=X,Y,
\label{eq:S_Wannier_expansion}
\end{equation}
where ${\boldsymbol\tau}_a$ specifies the position of sector $a$
inside the unit cell. Retaining the density-density part of the
interaction gives
\begin{equation}
H_C
=
\frac{1}{2}
\sum_{{\bf R},{\bf R}'}
\sum_{ab}
V_{ab}({\bf R}-{\bf R}')
n_{{\bf R}a}
n_{{\bf R}'b},\quad V_{ab}({\bf R}-{\bf R}')
=
\int d{\bf r}\,d{\bf r}'\,\left|
w_a({\bf r}-{\bf R}-{\boldsymbol\tau}_a)
\right|^2
W({\bf r}-{\bf r}')
\left|
w_b({\bf r}'-{\bf R}'-{\boldsymbol\tau}_b)
\right|^2.
\label{eq:S_Wannier_interaction}
\end{equation}
Equation~(\ref{eq:S_Wannier_interaction}) makes explicit that the effective
interaction between the low-energy sectors is a matrix element of the
screened Coulomb interaction weighted by their real-space
orbital-sublattice wave functions. Fourier transforming Eq.~(\ref{eq:S_Wannier_interaction}) gives
\begin{equation}
H_{\rm int}
=
\frac{1}{2}
\sum_{\bf q}
\sum_{ab}
V_{ab}({\bf q})
\rho_a({\bf q})
\rho_b(-{\bf q}),
\label{eq:S_bare_interaction}
\end{equation}
where
\begin{align}
V_{ab}({\bf q})
=
\sum_{\bf R}e^{-i{\bf q}\cdot{\bf R}}
\int d{\bf r}\,d{\bf r}'\,
&
|w_a({\bf r}-{\boldsymbol\tau}_a)|^2
W({\bf r}-{\bf r}'-{\bf R})
|w_b({\bf r}'-{\boldsymbol\tau}_b)|^2.
\label{eq:S_Vab_integral}
\end{align}
We define 
$V_{\parallel}({\bf q})
\equiv
V_{XX}({\bf q})
\simeq
V_{YY}({\bf q})$, 
and 
$V_{\perp}({\bf q})
\equiv
V_{XY}({\bf q})
\simeq
V_{YX}({\bf q})$. The former describes interactions within the dominant
orbital-sublattice sectors, whereas the latter connects the two
distinct sectors. Because the $X$ and $Y$ sectors correspond to different magnetic
sublattices and orthogonal $d_{xz}$ and $d_{yz}$ orbital characters,
their interaction matrix element can be substantially
reduced compared with the corresponding intra-sector matrix element.
The spatial displacement between the two sublattices as well as the distinct orbital form factors suppress the inter-sector
component. We therefore consider the regime $|V_{\perp}({\bf q})|
\ll |{V_{\parallel}({\bf q})}|$.

The orbital-sublattice density operator entering
Eq.~(\ref{eq:S_bare_interaction}) is
\begin{equation}
\rho_a({\bf q})
=
\sum_{{\bf k},s}
d^{\dagger}_{{\bf k}+{\bf q},as}
d_{{\bf k},as}.
\label{eq:S_density}
\end{equation}
Projecting the orbital operator onto the relevant low-energy
altermagnetic band, 
$d_{{\bf k}as}
=
u_{as}({\bf k})c_{{\bf k}s}$, gives
\begin{equation}
\rho_a({\bf q})
=
\sum_{{\bf k},s}
F_{as}({\bf k},{\bf q})
c^{\dagger}_{{\bf k}+{\bf q},s}
c_{{\bf k}s},
\label{eq:S_projected_density}
\end{equation}
where 
$F_{as}({\bf k},{\bf q})
=
u_{as}^{*}({\bf k}+{\bf q})
u_{as}({\bf k})$ is the orbital-sublattice density form factor. Substituting Eq.~(\ref{eq:S_projected_density}) into
Eq.~(\ref{eq:S_bare_interaction}) yields
\begin{align}
H_{\rm int}
=
\frac{1}{2}
\sum_{{\bf k},{\bf k}',{\bf q}}
\sum_{ss'}
&
\Gamma_{ss'}({\bf k},{\bf k}';{\bf q})
c^{\dagger}_{{\bf k}+{\bf q},s}
c_{{\bf k}s}c^{\dagger}_{{\bf k}'-{\bf q},s'}
c_{{\bf k}'s'},
\label{eq:S_projected_interaction}
\end{align}
with the projected interaction vertex
\begin{equation}
\Gamma_{ss'}({\bf k},{\bf k}';{\bf q})
=
\sum_{ab}
V_{ab}({\bf q})
F_{as}({\bf k},{\bf q})
F_{bs'}({\bf k}',-{\bf q}).
\label{eq:S_Gamma}
\end{equation}
The effective interaction $U_{ss'}({\bf Q})$ for a $s$-wave density-wave gap opening is obtained by averaging this projected vertex over the nested
portions of the Fermi surface, 
$U_{ss'}({\bf Q})
=
\left\langle
\Gamma_{ss'}({\bf k},{\bf k}';{\bf Q})
\right\rangle_{\rm nest}$.  We next show explicitly how the strong spin-orbital-sublattice
locking leads to a smaller inter-spin interaction.
For simplicity, we denote by $\eta\ll1$ the characteristic minority
orbital-sublattice amplitude on the nested Fermi-surface regions,
$\eta_{\bf k}\sim\eta$. From
Eq.~(\ref{eq:S_wavefunctions}), the dominant orbital weights satisfy
\begin{align}
|u_{X\uparrow}|^2
&=
1-O(\eta^2),
&
|u_{Y\uparrow}|^2
&=
O(\eta^2),
\nonumber\\
|u_{Y\downarrow}|^2
&=
1-O(\eta^2),
&
|u_{X\downarrow}|^2
&=
O(\eta^2),
\label{eq:S_orbital_weights}
\end{align}
and then, the corresponding density
form factors scale as
\begin{align}
F_{X\uparrow}({\bf Q})&\sim O(1),
&
F_{Y\uparrow}({\bf Q})&\sim O(\eta^2),
\nonumber\\
F_{Y\downarrow}({\bf Q})&\sim O(1),
&
F_{X\downarrow}({\bf Q})&\sim O(\eta^2).
\label{eq:S_formfactor_scaling}
\end{align}
The leading intra-spin interaction is therefore
\begin{equation}
U_{\uparrow\uparrow}({\bf Q})
=
V_{XX}({\bf Q})
|F_{X\uparrow}({\bf Q})|^2
+
O(\eta^2 V_{\perp})
+
O(\eta^4 V_{\parallel}),\quad
U_{\downarrow\downarrow}({\bf Q})
=
V_{YY}({\bf Q})
|F_{Y\downarrow}({\bf Q})|^2
+
O(\eta^2 V_{\perp})
+
O(\eta^4 V_{\parallel}).
\end{equation}
The altermagnetic crystalline symmetry relating the two sectors then
gives
\begin{equation}
U_{\uparrow\uparrow}({\bf Q})
\simeq
U_{\downarrow\downarrow}({\bf Q})
\equiv
U_{\parallel}({\bf Q}).
\label{eq:S_Uparallel}
\end{equation}
The inter-spin channel has a qualitatively different structure:
\begin{align}
U_{\uparrow\downarrow}({\bf Q})
\simeq&
\;
V_{XY}({\bf Q})
F_{X\uparrow}({\bf Q})
F_{Y\downarrow}^{*}({\bf Q})+
V_{XX}({\bf Q})
F_{X\uparrow}({\bf Q})
F_{X\downarrow}^{*}({\bf Q})+
V_{YY}({\bf Q})
F_{Y\uparrow}({\bf Q})
F_{Y\downarrow}^{*}({\bf Q})
+\cdots .
\label{eq:S_Uupdown_expansion}
\end{align}
The first term directly couples the two distinct dominant
orbital-sublattice sectors and is therefore controlled by the
finite-${\bf Q}$ inter-sector matrix element $V_{\perp}({\bf Q})$.
The second and third terms require the minority orbital-sublattice
components of the altermagnetic wave functions and are suppressed by
the small mixing parameter $\eta$. Consequently,
\begin{equation}
\frac{
U_{\uparrow\downarrow}({\bf Q})
}{
U_{\uparrow\uparrow}({\bf Q})
}\sim
O\left(
\frac{V_{XY}({\bf Q})}
{V_{XX}({\bf Q})}
\right)
+
O(\eta^2).
\label{eq:S_ratio}
\end{equation}
Therefore, in the regime of strong spin-orbital-sublattice locking,
$\eta\ll1$, with a weak finite-${\bf Q}$ interaction matrix
element between the two distinct low-energy sectors, 
$|V_{XY}({\bf Q})|
\ll
|V_{XX}({\bf Q})|$, one obtains
\begin{equation}
U_{\uparrow\downarrow}({\bf Q})
\ll
U_{\uparrow\uparrow}({\bf Q})
\simeq
U_{\downarrow\downarrow}({\bf Q}).
\label{eq:S_final_hierarchy}
\end{equation}

The above hierarchy provides a microscopic basis for treating the
two spin-resolved density-wave instabilities as approximately
independent at leading order. The dominant interaction in each sector
is
\begin{equation}
U_{\uparrow\uparrow}({\bf Q})
\simeq
U_{\downarrow\downarrow}({\bf Q})
\equiv
U_{\parallel}({\bf Q}).
\end{equation}
As the two sectors are related by altermagnetic crystalline
symmetry, they develop comparable density-wave amplitudes, $|\Delta_{\uparrow}|\simeq|\Delta_{\downarrow}|$, while their
density-wave modulations reside predominantly in different
spin-orbital sectors. The much smaller
$U_{\uparrow\downarrow}({\bf Q})$ produces only a subleading coupling
between the two condensates, and does not qualitatively
modify the magnitude of the individual
density-wave gaps, although it can contribute to the energetics of
the relative phase between $\Delta_{\uparrow}$ and
$\Delta_{\downarrow}$.

\section{Derivation of phase fluctuations}

In this section, we derive the contribution of phase fluctuations. We first clarify their physical importance: phase fluctuations play a crucial role in density-wave systems, as they can suppress long-range order through thermal averaging even when the amplitude of the order parameter remains finite. Specifically, following Eq.~(\ref{drho}) as well as Eq.~(\ref{dgap}) with $\phi(R)=\phi_e+\delta\phi(R)$, the charge-density modulation after electronic condensation is written as \begin{equation} \rho({R})=\rho_{Q}e^{i{Q}\cdot{R}}+c.c.=4V^{-1}|\Delta|\cos({Q}\cdot{R}+\phi_e+\delta\phi). \end{equation} 

Using the statistic average $\langle\delta\phi\rangle=0$ and $\langle\delta\phi^2\rangle\ne0$, the thermal average can be evaluated via the cumulant expansion~\cite{mahan2013many,abrikosov2012methods,peskin2018introduction}, 
\begin{align}\label{DW} &\langle\cos(\phi)\rangle={\rm Re}[e^{i\phi_e}\langle{e^{i\delta\phi}}\rangle]={\rm Re}\Big[e^{i\phi_e}\sum_{n=0}\frac{\langle(i\delta\phi)^{2n}\rangle}{(2n)!}\Big]={\rm Re}\Big[e^{i\phi_e}\sum_{n=0}\frac{(2n-1)!!(-1)^n\langle\delta\phi^{2}\rangle^n}{(2n)!}\Big]={\rm Re}\Big[e^{i\phi_e}\sum_{n=0}\frac{(-1)^n\langle\delta\phi^{2}\rangle^n}{2^nn!}\Big]\nonumber\\
&={\rm Re}\Big[e^{i\phi_e}e^{-\langle{\delta\phi}^2\rangle/2}\Big]=\cos(\phi_e)e^{-\langle{\delta\phi}^2\rangle/2}.
\end{align}
As a result, the thermally averaged density-wave order takes the form
\begin{eqnarray}
\langle\rho({R})\rangle=4V^{-1}|\Delta|\exp(-\langle{\delta\phi}^2\rangle/2)\cos({Q}\cdot{R}+\phi_e),
\end{eqnarray}
where a Debye–Waller–like factor~\cite{kittel1963quantum,sakurai2020modern} $\exp({-\langle{\delta\phi^2}\rangle}/2)$ naturally emerges from thermal phase fluctuations.  Then, the temperature dependence of the density-wave order is governed not only by the gap amplitude, but also by phase fluctuations. 

The derived expression of phase-fluctuating density-wave order here is mathematically rigorous and incorporates phase fluctuations to all orders through Gaussian contraction~\cite{peskin2018introduction}.  Thus, only in the limit of weak phase fluctuations, where long-range phase coherence is preserved, can the density-wave modulation induced by the underlying CDW condensate amplitude $|\Delta_s|$ be directly observed. By contrast, strong thermal phase fluctuations can destroy the observable density-wave modulation $\langle\rho_s({r})\rangle$ even though the underlying gap amplitude $|\Delta_s|$ (an associated pseudogap in the electronic spectrum) remains finite. It is precisely this separation between phase coherence and condensate amplitude that accounts for the suppression of long-range coherence at finite temperatures, even in the presence of a robust single-particle excitation gap (measured in STM or ARPES) and naturally  gives rise to two distinct temperature scales: $T_c$ marks the loss of static  density-wave order $\rho_s({r})$ due to strong phase fluctuations, whereas $T_g$ marks the disappearance of the underlying condensate amplitude $|\Delta_s|$ and the associated pseudogap. The persistence of the gap above $T_c$ is therefore a defining signature of a fluctuation-driven pseudogap regime.

\subsection{Phase pinning by impurities}

In realistic systems, impurities~\cite{PhysRevB.17.535} or lattice imperfections~\cite{gruner1985charge,gruner1988dynamics} can pin the density-wave by introducing an excitation gap (i.e., an effective mass) in the phase mode. This excitation gap regularizes the infrared divergence of phase correlations that underlies the Mermin–Wagner theorem~\cite{hohenberg1967existence,mermin1966absence,coleman1973there}. Specifically, as proposed and developed by Fukuyama and Lee~\cite{PhysRevB.17.535} as well as by Rice, Whitehouse, and Littlewood~\cite{PhysRevB.24.2751}, the interaction between the density-wave and an impurity potential $U_i({R}-{R}_i)$ located at ${R}_i$ is written as
\begin{equation}
H_i=-\sum_i\int{d{R}}\,\rho({R})U_i({R}-{R}_i).  
\end{equation}
Assuming a short-range impurity potential $U_i({R}-{R}_i)=u\delta({R}-{R}_i)$ with $u>0$, the action associated with impurity pinning of the density-wave phase becomes
\begin{equation}
  S_i=\int{dR}\Big\{4uV^{-1}|\Delta|\sum_i\cos[{Q}{\cdot}{R}_i+\phi_e({R}_i)+\delta\phi(R)]\Big\},  \label{SSI}
\end{equation}
which implies that the equilibrium phase configuration $\phi_e$ is no longer spatially uniform, but instead becomes inhomogeneous due to impurity pinning.  In the dilute impurity limit, the density-wave locally distorts to maximize its coupling to the impurity potential, leading to the condition~\cite{PhysRevB.17.535,PhysRevB.24.2751}
\begin{equation}\label{dic}
 \langle\cos[{Q}{\cdot}{R}_i+\phi_e({R}_i)]\rangle\approx-1.
\end{equation}
As will be shown below, this impurity-induced pinning effectively generates an excitation gap in the phason spectrum, which regularizes
the infrared divergence of the phase correlations~\cite{auerbach2012interacting} underlying
the Mermin-Wagner theorem~\cite{hohenberg1967existence,mermin1966absence,coleman1973there} that forbids long-range ordering in low-dimensional systems.

\subsection{Dynamics of phase fluctuations}  

Then, we derive the dynamics of the phase fluctuations in the presence of impurities.
Starting from Eqs.~(\ref{ace}) and (\ref{SSI}), the Euler–Lagrange equation with respect to the phase fluctuation field $\delta\phi$ reads~\cite{peskin2018introduction}
\begin{equation}
\partial_{\mu}\left[\frac{\partial (S^e_{\rm eff}+S_i)}{\partial(\partial_{\mu}\delta\phi/2)}\right]
=\frac{\partial (S^e_{\rm eff}+S_i)}{\partial(\delta\phi/2)} ,
\end{equation}
which leads to the equation of motion
\begin{equation}
\Big[D\partial_t^2-D f_c({v}_{Q/2}\cdot\partial_{R})^2\Big]\frac{\delta\phi(R)}{2}
-8uV^{-1}|\Delta|
\sum_i \sin\big[{Q}\cdot{R}_i+\phi_e({R}_i)+\delta\phi(R)\big]=0 .
\end{equation}
To proceed, we treat the phase fluctuations within a field-theoretical cumulant expansion,
retaining only a single phase operator while contracting the remaining fields.
Using the trigonometric identity
\begin{align}
\sin(\phi_e+\delta\phi)
=\sin\phi_e\cos\delta\phi+\cos\phi_e\sin\delta\phi ,
\end{align}
and assuming $\langle\delta\phi\rangle=0$, the linear response of the impurity term is governed by
the second contribution. The sine of the phase fluctuation is expanded as
\begin{align}
\sin\delta\phi(R)
=\sum_{n=0}^{\infty}\frac{(-1)^n}{(2n+1)!}\,\delta\phi^{\,2n+1}(R).
\end{align}
Within the cumulant approximation~\cite{peskin2018introduction,mahan2013many,abrikosov2012methods}, we retain only a single uncontracted phase operator
while contracting the remaining $2n$ fields according to Wick's theorem~\cite{peskin2018introduction},
\begin{align}
\delta\phi^{\,2n+1}(R)\;\rightarrow\;(2n-1)!!\,\langle\delta\phi^2(R)\rangle^{n}\,\delta\phi(R).
\end{align}
Substituting this back, we obtain
\begin{align}
\sin\delta\phi(R)
&\rightarrow \sum_{n=0}^{\infty}\frac{(-1)^n}{(2n+1)!}\,(2n-1)!!\,\langle\delta\phi^2(R)\rangle^{n}\,\delta\phi(R)=\sum_{n=0}^{\infty}\frac{(-1)^n}{2^n n!}\,\langle\delta\phi^2(R)\rangle^{n}\,\delta\phi(R)
=\exp\!\Big[-\frac{1}{2}\langle\delta\phi^2\rangle\Big]\delta\phi(R).
\end{align}
Therefore, the impurity term in the linear-response level becomes
\begin{align}
\sin(\phi_e+\delta\phi)
\;\rightarrow\;
\cos\phi_e\,\exp\!\Big[-\frac{1}{2}\langle\delta\phi^2\rangle\Big]\delta\phi(R).
\end{align}
and then, the equation of motion of the phase fluctuations reads
\begin{equation}
\Bigg\{D\partial_t^2-Df_c({v_{Q/2}}\cdot{\partial_{R}})^2-16uV^{-1}|\Delta|\exp\!\Big[-\frac{1}{2}\langle\delta\phi^2\rangle\Big]\sum_i\cos[{Q}{\cdot}{R}_i+\phi_e({R}_i)]\Bigg\}\frac{\delta\phi(R)}{2}=0.  
\end{equation}

Now, substituting the dilute-impurity condition~\cite{PhysRevB.17.535,PhysRevB.24.2751} of Eq.~(\ref{dic}), one finally reaches
\begin{equation}\label{EOM}
\Bigg\{D\partial_t^2-Df_c({v_{Q/2}}\cdot{\partial_{R}})^2+16un_iV^{-1}|\Delta|\exp\!\Big[-\frac{1}{2}\langle\delta\phi^2\rangle\Big]\Bigg\}\frac{\delta\phi(R)}{2}=0.  
\end{equation}
From this equation, the phason energy spectrum follows as
\begin{equation}
  \Omega^2({q})=16un_i(DV)^{-1}|\Delta|\exp\!\Big[-\frac{1}{2}\langle\delta\phi^2\rangle\Big]+f_c({v_{Q/2}}\cdot{q})^2,
\end{equation}
or equivalently, 
\begin{equation}\label{phasonenergy}
  \Omega^2({q})=m^2_P(T)+f_c({v_{Q/2}}\cdot{q})^2,\quad\quad\quad\quad m^2_P(T)=16un_i(DV)^{-1}|\Delta|\exp\!\Big[-\frac{1}{2}\langle\delta\phi^2\rangle\Big].
\end{equation}
Clearly, the impurity-induced pinning effectively generates a finite excitation gap (effective phason mass) $m_p(T)$ in the phason spectrum. The pinning strength therefore acquires a self-consistent temperature dependence,
\begin{equation}\label{mpEq}
m_p^2(T)=m_p^2(0)\exp\!\Big[-\frac{1}{2}\langle\delta\phi^2(T)\rangle\Big]\,
\frac{|\Delta(T)|}{|\Delta(0)|},
\end{equation}
which explicitly captures the feedback between thermal phase fluctuations and the effective pinning potential:
increasing temperature enhances $\langle\delta\phi^2(T)\rangle$ and reduces $|\Delta(T)|$, both of which suppress the
coherent impurity pinning and lead to a progressive softening of the phason mode upon warming.
To the best of our knowledge, although such a temperature-dependent pinning form has been widely employed at the phenomenological level in the density-wave literature~\cite{PhysRevB.24.2751,maki1986thermal}, a controlled microscopic derivation starting from an effective action with disorder and phase fluctuations has been lacking. The present work fills this gap by providing a field-theoretical derivation based on the cumulant  treatment of phase fluctuations.

\subsection{Statistic average of phase fluctuations}

From the equation of motion of the phase fluctuations in Eq.~(\ref{EOM}),
the average of thermal phase fluctuations can be obtained within quantum-statistic mechanism~\cite{abrikosov2012methods,mahan2013many} either via the fluctuation-dissipation theorem
or within the Matsubara formalism.
The two approaches are fully equivalent and yield identical results, reflecting the bosonic description of the phase degree of freedom as a collective phason (Nambu-Goldstone~\cite{nambu1960quasi,nambu2009nobel,goldstone1961field,goldstone1962broken}) mode.\\

{\sl Fluctuation dissipation theorem.--}Considering the thermal fluctuations, the dynamics of the phase from Eq.~(\ref{EOM}) is given by
\begin{equation}
\big[D\Omega^2({q})-\omega^2+iD\omega{\gamma}\big]\frac{\delta\phi({\omega,{q}})}{2}={J}_{\rm th}(\omega,{q}).    
\end{equation}  
Here, we have introduced a thermal field ${J}_{\rm th}(\omega,{q})$ that obeys the fluctuation-dissipation theorem~\cite{Landaubook}:
\begin{equation}
\langle{J}_{\rm th}(\omega,{q}){J}^*_{\rm th}(\omega',{q}')\rangle=\frac{(2\pi)^3D\gamma\omega\delta(\omega-\omega')\delta({q-q'})}{\tanh(\beta\omega/2)},   
\end{equation}
and $\gamma=0^+$ is a phenomenological damping constant. From this dynamics, the average of the phase fluctuations is given by
\begin{eqnarray}
  \frac{1}{4}\langle{\delta\phi^2}\rangle&=&\int\frac{d\omega{d\omega'}d{q}d{q'}}{(2\pi)^6}\frac{\langle{J}_{\rm th}(\omega,{q}){J}^*_{\rm th}(\omega',{q}')\rangle}{D^2\big[\big(\omega^2\!-\!\Omega^2({q})\big)\!-\!i\omega\gamma\big]\big[({\omega'}^2\!-\!\Omega^2({q}')\big)\!+\!i\omega'\gamma\big]}=\int\frac{d\omega{d{q}}}{D(2\pi)^2}\frac{\gamma\omega/\tanh(\beta\omega/2)}{\big[\omega^2\!-\!\Omega^2({q})\big]^2+\omega^2\gamma^2}\nonumber\\
&=&\int\frac{{d{q}}}{(2\pi)^2}\frac{[2n_B\big(\Omega({q})\big)+1]}{2D\Omega({q})}.\label{EEEE}
\end{eqnarray}

{\sl Matsubara formalism.--}Within the Matsubara formalism, by mapping into the imaginary-time space, from Eq.~(\ref{EOM}),  the thermal phase fluctuation reads~\cite{abrikosov2012methods}
\begin{eqnarray}
   \frac{1}{4}\langle{\delta\phi^2}\rangle&\!=\!&\int\frac{d{q}}{(2\pi)^2}\Big[\Big\langle\Big|\frac{\delta\phi^*(\tau,{q})}{2}\frac{\delta\phi(\tau,{q})}{2}e^{-\int^{\beta}_{0}d\tau{d{q}}D{\delta\phi^*(\tau,{q})}(\Omega^2({q})-\partial_{\tau}^2){\delta\phi(\tau,{q})}/4}\Big|\Big\rangle\Big]\nonumber\\
  &\!=\!&\int\frac{d{q}}{(2\pi)^2}\Big[\frac{1}{\mathcal{Z}_0}{\int}D\delta\phi{D\delta\phi^*}\frac{\delta\phi^*(\tau,{q})}{2}\frac{\delta\phi(\tau,{q})}{2}e^{-\int^{\beta}_{0}d\tau{d{q}}D{\delta\phi^*(\tau,{q})}(\Omega^2({q})-\partial_{\tau}^2){\delta\phi(\tau,{q})}/4}\Big]\nonumber\\
   &=&\!\!\!\int\frac{d{q}}{(2\pi)^2}\frac{1}{\mathcal{Z}_0}{\int}D\delta\phi{D\delta\phi^*}\delta_{J^*_{q}}\delta_{J_{q}}e^{-\int^{\beta}_{0}d\tau{d{q}}[D{\delta\phi^*(\tau,{q})}(\Omega^2({q})-\partial_{\tau}^2){\delta\phi(\tau,{q})}/4+J_{q}\delta\phi({q})/2+J^*_{q}\delta\phi^*({q})/2]}\Big|_{J=J^*=0}\nonumber\\
  &=&\!\!\!\int\frac{d{q}}{(2\pi)^2}\frac{1}{D}\delta_{J^*_{q}}\delta_{J_{q}}\exp\Big\{-\int_0^{\beta}{d\tau}\sum_{q'}J_{q'}\frac{1}{\partial_{\tau}^2-\Omega^2({q})}J^*_{q'}\Big\}\Big|_{J=J^*=0}=-\int\frac{d{q}}{(2\pi)^2}\frac{1}{\beta}\sum_{\omega_n}\frac{1}{D}\frac{1}{(i\Omega_n)^2\!-\!\Omega^2({q})}\nonumber\\
 &=&\int\frac{{d{q}}}{(2\pi)^2}\frac{2n_B(\Omega({q}))+1}{2D\Omega({q})},\label{mf}
\end{eqnarray}
which is exactly the same as the one in Eq.~(\ref{EEEE}) obtained via the fluctuation dissipation theorem. Here, $\Omega_n=2n\pi{T}$ represents the Bosonic Matsubara frequencies; ${J_{q}}$ denotes the generating functional and $\delta{J_{q}}$ stands for the functional derivative.\\

\section{Numerical analysis}
 
\subsection{Simulation treatments}

Our simulation consists of two complementary steps, which together establish a self-consistent connection between the microscopic band structure and the finite-momentum ordered state.

First, we construct an effective tight-binding model by fitting to the underlying  band structure obtained from our DFT calculation (Fig.~1 in the main text).
For each spin component $s=\pm$, the electronic dispersion is taken as~\cite{PhysRevX.12.011028} 
\begin{equation}
\xi_{{k},s}
=-2t(\cos k_x+\cos k_y)
-4t'\cos k_x \cos k_y
-2t_j s_z(\cos k_x-\cos k_y)
-\mu ,
\end{equation}
where $t$ and $t'$ denote the nearest- and next-nearest-neighbor hopping amplitudes, $t_j$ parameterizes the spin-dependent anisotropic term, and $\mu$ is the chemical potential. We substitute the ordering wave vector ${Q}=(\pi,\pi)$ into the gap equation and evaluate the order-parameter magnitude
$|\Delta(T)|$ using the fitted tight-binding band structure.
This procedure yields the full temperature dependence of the density-wave gap and thereby determines
all gap-related quantities in a self-consistent manner.

\begin{table}[htb]
\centering
  \caption{Specific parameters used in our numerical calculation. The tight-binding parameters entering the dispersion are obtained by fitting the
DFT band structure and are kept fixed throughout the analysis in the main text.  The effective interaction strength $V$ is fixed by fitting the low-temperature gap magnitude
$|\Delta(T=20~\text{K})|=35~$meV measured experimentally~\cite{Jiang2025}.
  From a theoretical perspective, the main physical conclusions reported here are robust against
moderate variations of the pinning strength $m_P(0)$ (see Sec.~\ref{secdwov}). For numerical implementation, a finite momentum cutoff $\Lambda$ is introduced in the momentum integral involving the Bose distribution function.}  
\renewcommand\arraystretch{1.5}   \label{Table}
\begin{tabular}{l l l l l l l l}
    \hline
    \hline
    tight-binding model parameters (meV)&\quad\quad\quad~$t$&\quad\quad\quad~$t'$&\quad\quad\quad~$t_j$&\quad\quad\quad~$\mu$\\   
    &\quad\quad90.375&\quad\quad$-36.9375$&\quad\quad~$194.24$&\quad\quad~$75.25$\\
     \hline
    model parameter used in gap equation&\quad\quad\quad~$|\Delta_s(T=20~\text{K})|$\\   &\quad\quad\quad\quad$35~$meV\\
     \hline
    model parameters used in phase fluctuations&\quad\quad$m_{P}(T=0)$\\   
    &\quad\quad20~meV\\
    \hline
    \hline
\end{tabular}\\
\end{table}

The resulting $|\Delta(T)|$ is then used as inputs for the temporal
stiffness $D$ [Eq.~(\ref{DEq})] and the condensation fraction $f_c(T)$ [Eq.~(\ref{fcEq})].
With these quantities, we self-consistently evaluate the impurity-induced pinning strength
[Eq.~(\ref{mpEq})] together with the thermal average of the phase fluctuations $\langle\delta\phi^2\rangle$. Owing to the Debye-Waller-like exponential structure of the phase-fluctuation renormalization,
all zero-point (quantum) fluctuation contributions can be absorbed into the zero-temperature
parameters $m_p(0)$ and $\rho_{\rm CDW}(0)$.
Throughout the numerical analysis, we therefore explicitly account only for the
thermally excited phase fluctuations,
\begin{equation}
\langle\delta\phi^2_{\rm th}\rangle/2=\int{\frac{d{q}}{(2\pi)^2}}\frac{2n_B(\Omega({q}))}{D\Omega({q})},
\end{equation}
where $\Omega({q})=\sqrt{m_p^2(T)+f_c(T)\,({v}_{Q/2}\!\cdot{q})^2}$. While the phason dispersion here is highly anisotropic, the phase-fluctuation field itself remains a 2D bosonic mode. All fluctuation contributions are evaluated through a 2D momentum-space integration.

The specific parameter values employed in the simulations, as well as the criteria used for
their selection and experimental fitting, are summarized in Table~\ref{Table}. The effective interaction strength $V$ can not be  fixed by the microscopic tight-binding band structure or DFT calculation, and therefore need to be determined by comparison with experiment. 
The inverse coupling $1/V$ is fixed to reproduce the experimentally observed low-temperature gap magnitude 
$|\Delta(T=20~\text{K})|\approx35~$meV~\cite{Jiang2025}.

\subsection{Calculated saddle-shaped NMR line profile arising from stripe spin-density modulation}

We consider that the local spin density varies periodically in real space as 
$M({\bf r}) = M_0 \cos({\bf Q}\cdot{\bf r}+\phi)$, where $M_0$ is the amplitude of the spin-density wave, ${\bf Q}$ is the stripe ordering wave vector, and $\phi$ is an arbitrary phase. The spatially modulated spin polarization produces a position-dependent hyperfine field 
$B_{\rm hf}({\bf r}) = A_{\rm hf} M({\bf r})$, with $A_{\rm hf}$ the hyperfine coupling constant. The corresponding local NMR resonance frequency is therefore
\begin{equation}
\nu({\bf r})
=
\nu_0+\gamma_n B_{\rm hf}({\bf r})
=
\nu_0+\Delta\nu \cos({\bf Q}\cdot{\bf r}+\phi),
\end{equation}
where $
\Delta\nu=\gamma_n A_{\rm hf}M_0$
 sets the maximum frequency shift induced by the stripe spin modulation.

The measured NMR spectrum corresponds to the probability distribution of local resonance frequencies throughout the sample. For an incommensurate stripe modulation, the phase 
$\theta={\bf Q}\cdot{\bf r}+\phi$
 can be regarded as uniformly distributed over $[0,2\pi)$. The ideal NMR line shape can therefore be written as
\begin{equation}
I(\nu)
=
\frac{1}{2\pi}
\int_0^{2\pi}
d\theta\,
\delta
\left[
\nu-\nu_0-\Delta\nu\cos\theta
\right]
=
\frac{1}{
\pi
\sqrt{
\Delta\nu^2-(\nu-\nu_0)^2
}
},
\qquad
|\nu-\nu_0|<\Delta\nu.
\end{equation}

Thus, a simple sinusoidal stripe spin-density wave gives rise to a characteristic double-edge or saddle-shaped NMR spectrum. The enhanced spectral weight near 
$\nu=\nu_0\pm\Delta\nu$ originates from the extrema of the spin-density modulation, where the spatial derivative of the local hyperfine field vanishes.

In a realistic system, the singularities at the two edges are broadened by disorder, finite instrumental resolution, nuclear dipolar broadening, and spatial fluctuations of the stripe order. These effects can be phenomenologically incorporated by introducing a finite imaginary part $\Gamma$ into the local resonance frequency, corresponding to a finite lifetime of the nuclear-spin excitation, and then, the broadened NMR line shape can be written analytically as
\begin{equation}
I(\nu)
=
-\frac{1}{\pi}
{\rm Im}
\left[
\frac{1}
{
\sqrt{
(\nu-\nu_0+i\Gamma)^2-\Delta\nu^2
}
}
\right].
\end{equation}

\subsection{Density of states of a perfectly nested CDW with finite broadening}

For a perfectly nested CDW state, the electronic dispersions connected by the ordering wave vector ${\bf Q}$ satisfy 
$\xi_{{\bf k}+{\bf Q}}=-\xi_{\bf k}$. The quasiparticle spectrum is then 
$E_{\bf k}^{\pm}
=
\pm\sqrt{\xi_{\bf k}^2+|\Delta|^2}$.  The retarded Green's function is therefore
\begin{equation}
G^{R}({\bf k},\omega)
=
\frac{\omega+i\Gamma+\xi_{\bf k}}
{(\omega+i\Gamma)^2-\xi_{\bf k}^2-|\Delta|^2}.
\end{equation}
where $\Gamma$ is a phenomenological broadening that accounts for the finite lifetime of the electronic quasiparticles.  The density of states is obtained from
\begin{equation}
N(\omega)
=
-\frac{1}{\pi}
{\rm Im}
\sum_{\bf k}
G^{R}({\bf k},\omega).
\end{equation}
Assuming a constant normal-state density of states $N_0$ near the Fermi energy, the normalized density of states is
\begin{equation}
\frac{N(\omega)}{N_0}
=
{\rm Re}
\Big[
\frac{\omega+i\Gamma}
{\sqrt{(\omega+i\Gamma)^2-|\Delta|^2}}
\Big].
\end{equation}
In the pseudogap phase, $\Gamma$ can be relatively large due to strong scattering from the gapless phason fluctuations. Below $T_c$, the phason acquires a finite gap, which strongly suppresses the low-energy phase fluctuations and the associated quasiparticle scattering. Consequently, $\Gamma$ is substantially reduced below $T_c$, leading to a sharpening of the CDW spectral features and the emergence of more pronounced coherence-like peaks in the density of states.


\section{Discussion of critical behaviors}

In this part, we discuss the critical behaviors associated with the phase transitions in our framework.

\subsection{Density-wave ordering vanishing}
\label{secdwov}

We first consider the density-wave ordering transition. Strong phase fluctuations can suppress the
long-range density-wave order through thermal averaging even when the gap amplitude remains finite.
As a result, the density-wave order parameter vanishes at a lower temperature $T_c<T_g$,
prior to the closing of the density-wave gap. Within our framework, this suppression is quantitatively captured by the Debye-Waller-like factor $\exp({-\langle{\delta\phi^2}\rangle}/2)$ 
arising from thermal phase fluctuations, which exponentially reduces the coherent density modulation. Consequently, the loss of density-wave order at $T_c$ is driven by phase disordering rather than by
the collapse of the gap amplitude. The intermediate regime $T_c<T<T_g$ thus corresponds to a pseudogap phase, characterized by a finite
gap amplitude but the absence of long-range density-wave order.

Mathematically, the condition
$\exp[-\langle\delta\phi^2_{\rm th}(T_c)\rangle/2]\ll 1$
has two important consequences.

First, it is equivalent to
$\langle\delta\phi^2_{\rm th}(T_c)\rangle/2\gg 1$, i.e.,
\begin{equation}
\left.\frac{\langle\delta\phi^2_{\rm th}\rangle}{2}
=\int\frac{d{q}}{(2\pi)^2}
\frac{2n_B\!\big[\sqrt{f_c(T)}\,({v}_{Q/2}\!\cdot{q})\big]}
{D\sqrt{f_c(T)}\,({v}_{Q/2}\!\cdot{q})}
\right|_{T=T_c}
\gg 1 .
\end{equation}
Importantly, this condition is independent of the bare pinning scale $m_p(0)$, as also proposed in the literature~\cite{PhysRevB.24.2751,maki1986thermal}.
Therefore, from a theoretical perspective, the transition temperature $T_c$ is
controlled by intrinsic material parameters (such as the electronic structure
and phase stiffness) rather than by sample-dependent pinning strength.

Second, near $T_c$ where $m_p(T_c)\simeq 0$, the absence of an excitation gap in the
phason spectrum removes the infrared protection against long-wavelength fluctuations.
As a result, the thermal phase fluctuations become strongly enhanced and tend to diverge,
in accordance with the Mermin-Wagner theorem~\cite{hohenberg1967existence,mermin1966absence,coleman1973there}.
This leads to a detrimental feedback mechanism in the vicinity of $T_c$:
the suppression of the pinning-induced mass $m_p(T)$ enhances phase fluctuations,
which in turn further suppress $m_p(T)$ through the Debye-Waller-like  renormalization.
Consequently, the density-wave order is expected to be rapidly destroyed near $T_c$,
exhibiting a sharp suppression that resembles a weakly first-order transition.

From a quantum-statistical standpoint, low-dimensional ordering systems are particularly susceptible to thermal fluctuations of the Nambu–Goldstone (phason) mode~\cite{nambu1960quasi,nambu2009nobel,goldstone1961field,goldstone1962broken}, whose infrared divergence precludes spontaneous breaking of continuous symmetries at any finite temperature~\cite{auerbach2012interacting}. Consequently, in the absence of additional symmetry-breaking mechanisms, genuine long-range density-wave order can only emerge in the singular limit $T=0$, where the transition at $T\rightarrow0^+$ is intrinsically sharp. In realistic materials, however, disorder~\cite{PhysRevB.17.535}, lattice commensurability~\cite{gruner1985charge,gruner1988dynamics}, or structural imperfections~\cite{gruner1985charge,gruner1988dynamics} explicitly lock the density-wave phase, generating a finite gap in the phason spectrum. This pinning regularizes long-wavelength fluctuations and stabilizes density-wave order over a finite temperature range. As temperature increases, thermal activation progressively weakens the effective pinning, leading to a softening of the phason mode. When the pinning scale is sufficiently reduced, fluctuation effects (phason excitations) again become dominant and destabilize the ordered state. Within this perspective, explicit phase locking converts an otherwise $T=0^+$ instability into a transition at a finite critical temperature $T_c$.

\subsection{Gap closing}

We next turn to the gap-closing transition.
As mentioned above, once the effective interaction strength $V$ is fixed by fitting the
experimental zero-temperature gap magnitude $|\Delta_s(T=0)|$ (which sets the pairing strength in the density-wave channel),
the full temperature dependence of the density-wave gap $|\Delta(T)|$ and hence the gap-closing temperature $T_g$
are uniquely determined within the model.  Since the gap equation is treated at the mean-field level, the gap-closing transition is continuous and
exhibits the standard second-order critical behavior. In particular, $|\Delta(T)|$ vanishes continuously upon
approaching $T_g$ from below, consistent with a mean-field universality class.  Notably, in the present work $2|\Delta(T)|/(k_BT_g)\sim3.59$, consistent with the mean-field expectation~\cite{abrikosov2012methods,mahan2013many,schrieffer1964theory}.  It is worth emphasizing that the thermal average of the Doppler-shift contribution~\cite{yang2025efficient,yang2025preformed,yang2025tractable,PhysRevB.98.094507,yang21theory}, $\langle({v_{Q/2}}\cdot\partial_{R}\delta\phi/{2})^2\rangle\sim\langle(q^2_x\delta\phi_{q}/{2})^2\rangle$ is itself free from infrared divergences. As a result, this quantity does not exhibit the infrared pathology associated with the Mermin–Wagner theorem and therefore does not destabilize the mean-field gap equation.

This situation is in clear contrast to lattice-driven density-wave systems, where strong
mass renormalization substantially suppresses the phason energy/velocity according to~\cite{Lee1974conductivity,gruner1988dynamics,gruner1985charge} 
\begin{equation}
\Omega^2({q})\Rightarrow \frac{m}{m^*}\Omega^2({q}),\quad\quad\quad m/m^*\ll1, 
\end{equation}
and hence, amplifies the Doppler-shift
effect induced by phase fluctuations.  Only in this case, the enhanced feedback of phase fluctuations
onto the fermionic spectrum can qualitatively modify the gap equation and can potentially drive the gap-closing transition into a first-order one in the lattice-driven density-wave systems~\cite{yang2025microscopic}.

\section{First-principles calculation method and quasi-2D nature of the electronic structure}

The first-principles DFT calculations are performed with the projector-augmented-wave method implemented in the Vienna Ab initio Simulation Package (\textsc{vasp})~\cite{VASP1,VASP2}.
We use the Perdew-Burke-Ernzerhof type exchange-correlation functional~\cite{PBE} under the generalized gradient approximation.
The cutoff energy of the plane wave basis set is 600 eV.
The Brillouin zone is sampled by a Monkhorst-pack $k$-point grids with spacing less than 0.18~\AA$^{-1}$, i.e., $9 \times 9 \times 5$.
Atomic structure relaxations were converged to 1 $\mu$eV and 5 meV/\AA for energy and forces, respectively.
A tighter energy convergence is enforced under 0.1 $\mu$eV for the following electronic structure analyses.
The effective masses were extracted from parabola fits to the band dispersion dense line sampling along the $\Gamma-X$ and $\Gamma-Y$ directions using dense line-mode $k$-point sampling near $\Gamma$.

\begin{figure}[h]
    \centering
    \includegraphics[width=0.95\linewidth]{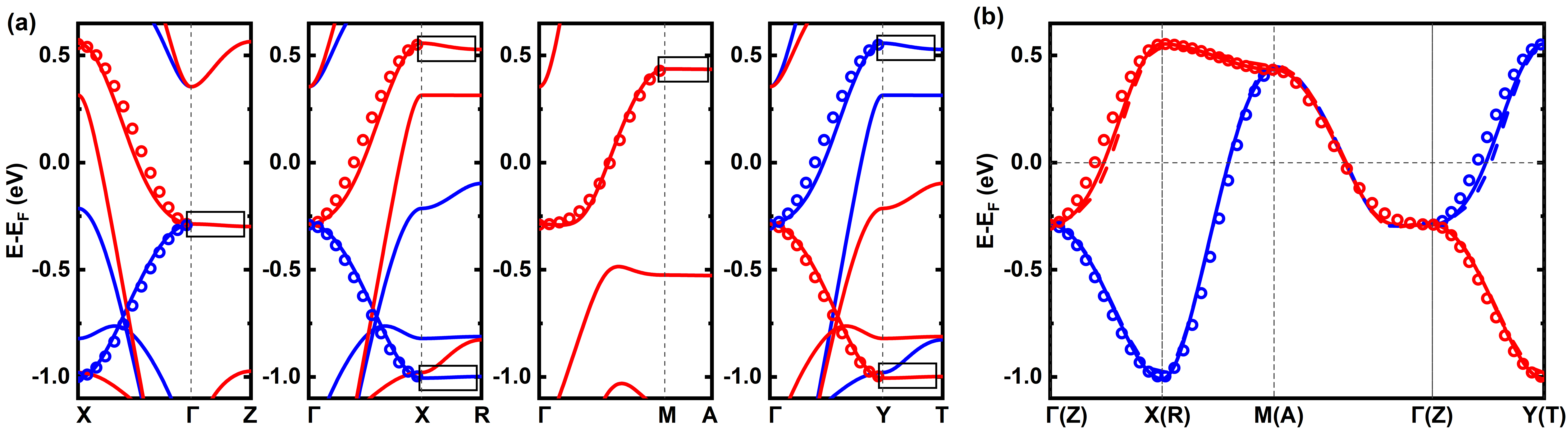}
   \caption{(a) Band structures along representative paths containing out-of-plane momentum segments, $X$-$\Gamma$-$Z$, $\Gamma$-$X$-$R$, $\Gamma$-$M$-$A$, and $\Gamma$-$Y$-$T$. The latter half of each path corresponds to the out-of-plane direction, namely $\Gamma$-$Z$, $X$-$R$, $M$-$A$, and $Y$-$T$, respectively. The bands highlighted by the boxes exhibit only negligible dispersion along these out-of-plane segments, indicating an extremely weak $k_z$ dependence of the electronic states relevant to the density-wave instability.
(b) Solid and dashed curves denote the selected low-energy DFT bands relevant to the density-wave instability at $k_z=0$ and $k_z=\pi$, respectively. Blue and red represent spin-up and spin-down channels, respectively; single-colored portions indicate spin-degenerate bands, while circles denote the corresponding tight-binding model bands. The very close similarity between the $k_z=0$ and $k_z=\pi$ dispersions, together with the nearly flat  out-of-plane dispersion, demonstrates the quasi-2D character of the relevant electronic structure and the negligible interlayer coupling of the electronic states relevant to the
density-wave instability.  The relevant low-energy electronic structure can then be effectively described by a 2D altermagnetic tight-binding model, and the altermagnetic spin splitting reconstructs the original 2D Fermi surface into spin-selective quasi-1D open Fermi-surface sheets.}
\label{figR1}
\end{figure}

One of the central treatments employed in the present work is the use of a 2D altermagnetic tight-binding model together with a 2D  fluctuation framework. Since KV$_2$Se$_2$O is a 3D bulk material, it is important to establish that the low-energy electronic states responsible for the density-wave instability are effectively confined to the crystal planes.

To address this issue, we performed  DFT calculations to examine the out-of-plane electronic dispersion. Figure~\ref{figR1}(a) shows the electronic band structures along several momentum paths containing substantial $k_z$ components, including $\Gamma$-$Z$, $X$-$R$, $M$-$A$, and $Y$-$T$. The relevant altermagnetic bands responsible for the observed density-wave
instability exhibit only negligible dispersion along these out-of-plane directions, as highlighted in the figure. This behavior indicates that the interlayer hopping associated with the electronic states relevant to the density-wave instability is extremely weak compared with the dominant in-plane electronic energy scales. To further quantify this observation, Fig.~\ref{figR1}(b) compares the relevant low-energy DFT band structures at $k_z=0$ and $k_z=\pi$. The two sets of dispersions are nearly indistinguishable throughout the energy window. The resulting negligible $k_z$ dependence demonstrates that the low-energy electronic structure possesses a pronounced quasi-2D  character. Consequently, the electronic states relevant to the density-wave instability can be accurately described by an effective 2D altermagnetic tight-binding model (circles in Fig.~\ref{figR1}), $\xi_{{k},s}
=-2t(\cos k_x+\cos k_y)
-4t'\cos k_x \cos k_y-2t_js_z(\cos k_x-\cos k_y)-\mu$. Starting from this effectively 2D electronic structure, the strong momentum-dependent spin splitting reconstructs the original Fermi surface into spin-selective open sheets with extended parallel segments. Therefore, the low-energy electronic structure is effectively quasi-2D, while the relevant nested Fermi-surface topology is spin-selective quasi-1D. Correspondingly, the phase-fluctuation theory developed in the present work remains intrinsically 2D, while the derived phason dispersion is highly anisotropic.

\begin{figure}[htb]
    \centering \includegraphics[width=0.85\linewidth]{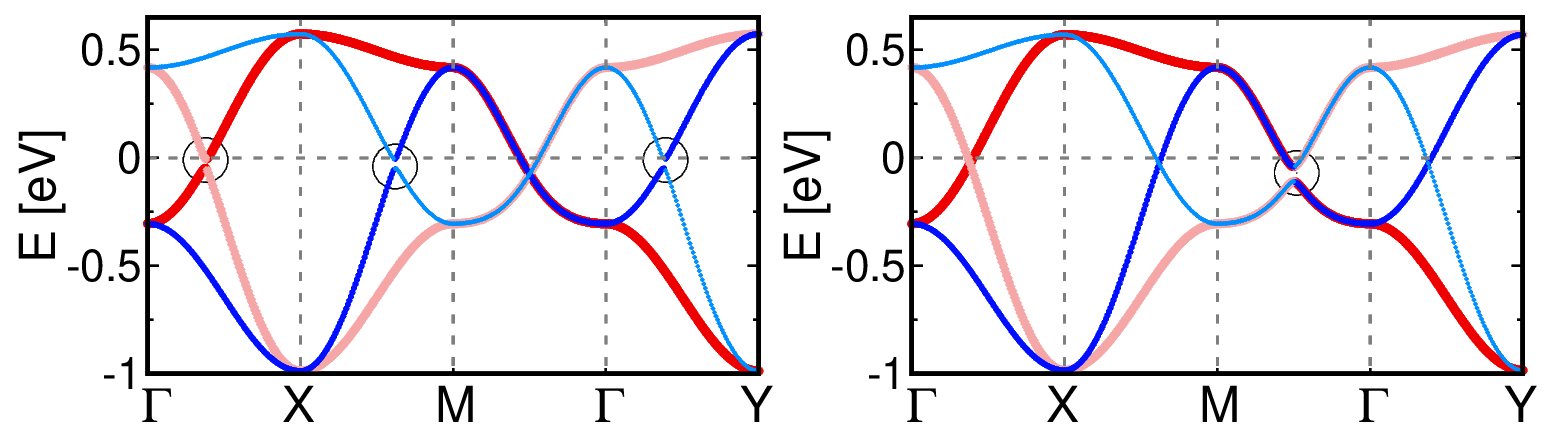}
\caption{{Calculated quasiparticle energy spectra in the density-wave state for momentum-dependent $d_{x^2-y^2}$-wave (left) and $d_{xy}$-wave (right) CDW order parameters. Red (blue) curves denote the spin-up (spin-down) quasiparticle bands. The darker thick curves represent the original electronic dispersion, while the lighter thin curves correspond to the backfolded bands induced by the density-wave order. The black dashed circles highlight the regions where the density-wave gap opens. For comparison, the gap magnitude is chosen to be approximately $35$ meV. A pronounced qualitative difference from the $s$-wave case can be clearly observed. For the $d_{x^2-y^2}$-wave CDW state, the gap opening along the $\Gamma$--$M$ direction is strongly suppressed due to the nodal structure of the form factor. In contrast, for the $d_{xy}$-wave CDW state, the gap opening along the $\Gamma$--$X$ and $\Gamma$--$Y$ directions is largely absent. These features are inconsistent with the experimentally observed gap opening along both the $\Gamma$--$M$ and $\Gamma$--$X/Y$ directions. Therefore, the spectroscopic data strongly disfavor a dominant $d$-wave interaction channel, whereas the $s$-wave spin-resolved CDW state naturally reproduces the observed momentum-space distribution of the gap.}}
\label{figD}
\end{figure}

\section{Exclusion of a $d$-wave CDW state}

In this section, we examine whether the experimentally observed CDW gap can be described by a $d$-wave order parameter $\Delta_{\bf k}$. The two commonly considered $d$-wave form factors are
\begin{equation}
\Delta_{\bf k}^{d_{x^2-y^2}}
=
\Delta_d(\cos k_x-\cos k_y),\quad
\Delta_{\bf k}^{d_{xy}}
=
\Delta_d\sin k_x\sin k_y.
\end{equation}

The corresponding gap-opening behavior, as shown in Fig.~\ref{figD}, is qualitatively different from the experimentally observed momentum dependence reported in spectroscopy. Experimentally, a finite gap opening is observed along both the $\Gamma$-$X/Y$ and $\Gamma$-$M$ directions~\cite{Jiang2025}. In particular, the simultaneous observation of finite gaps at $\theta_{\bf k}=0$ and $\theta_{\bf k}=\pi/4$ is incompatible with either $d$-wave form factor: the $d_{xy}$ state necessarily has a node at $\theta_{\bf k}=0$, whereas the $d_{x^2-y^2}$ state necessarily has a node at $\theta_{\bf k}=\pi/4$.

Importantly, these nodes are imposed by symmetry and therefore cannot be removed by increasing the magnitude of $\Delta_d$. A larger $\Delta_d$ only enhances the gap away from the nodal directions, while the gap remains exactly zero along the corresponding symmetry-enforced nodal directions. The experimentally observed finite gap opening along both $\Gamma$-$X/Y$ and $\Gamma$-$M$ therefore strongly disfavors a dominant $d$-wave CDW state. By contrast, the $s$-wave spin-resolved CDW state considered in our work has a nodeless form factor $\Delta_{\bf k}^{s}=\Delta_s$, which naturally produces a finite gap along both high-symmetry directions. The resulting gap-opening behavior is therefore fully consistent with the available spectroscopic measurements.

\bibliography{ref.bib}